\documentclass[11pt,english]{article}
\usepackage[latin9]{inputenc}
\usepackage{geometry}
\usepackage{xcolor}
\usepackage{array}
\usepackage{booktabs}
\usepackage{multirow}
\usepackage{amsmath}
\usepackage{amssymb}
\usepackage{graphicx}

\makeatletter

\providecommand{\tabularnewline}{\\}

\numberwithin{equation}{section}
\numberwithin{figure}{section}

\usepackage{amsmath,amsfonts,amssymb}
\usepackage{graphicx}
\usepackage{array}
\usepackage{wrapfig}
\usepackage{tikz}
\usepackage{rotating}
\usepackage{subcaption}

\usepackage{algorithm}
\usepackage{algorithmic}
\usepackage{setspace}
\usepackage{booktabs}
\usepackage{chngcntr}
\usepackage{listings}
\usepackage{color}

\definecolor{mygreen}{rgb}{0,0.6,0}
\definecolor{mygray}{rgb}{0.5,0.5,0.5}
\definecolor{mymauve}{rgb}{0.58,0,0.82}

\counterwithout{figure}{section}

\date{}

\def\E{\mathbb{E}}

\makeatother

\usepackage{babel}
\begin{document}
\title{Kernel Mean Embedding Based Hypothesis Tests for Comparing Spatial
Point Patterns}
\author{Raif M. Rustamov and James T. Klosowski, AT\&T Labs Research}
\maketitle
\begin{abstract}
This paper introduces an approach for detecting differences in the
first-order structures of spatial point patterns. The proposed approach
leverages the kernel mean embedding in a novel way by introducing
its approximate version tailored to spatial point processes. While
the original embedding is infinite-dimensional and implicit, our approximate
embedding is finite-dimensional and comes with explicit closed-form
formulas. With its help we reduce the pattern comparison problem to
the comparison of means in the Euclidean space. Hypothesis testing
is based on conducting $t$-tests on each dimension of the embedding
and combining the resulting $p$-values using one of the recently
introduced $p$-value combination techniques. If desired, corresponding
Bayes factors can be computed and averaged over all tests to quantify
the evidence against the null. The main advantages of the proposed
approach are that it can be applied to both single and replicated
pattern comparisons and that neither bootstrap nor permutation procedures
are needed to obtain or calibrate the $p$-values. Our experiments
show that the resulting tests are powerful and the $p$-values are
well-calibrated; two applications to real world data are presented.
\end{abstract}

\section{Introduction}

Comparison of spatial point patterns is of practical importance in
a number of scientific fields including ecology, epidemiology, and
criminology. For example, such comparisons may reveal differential
effects of the environment on plant species spread, uncover spatial
variation in disease risk, or detect seasonal differences in crime
locations (see e.g. \cite{spatstat}). While exploratory analyses
are vital for obtaining deep insights about pattern differences, such
analyses can be subjective unless supplemented with formal hypothesis
tests. 

In this paper we are interested in comparing the first-order structures
of point patterns. Consider two point processes $P$ and $Q$ over
the region $A\subset\mathbb{R}^{2}$ with the first-order intensities
given by $\lambda^{P}(\cdot$) and $\lambda^{Q}(\cdot)$. Given realizations
from these processes, we would like to detect whether there are statistically
significant differences in the first-order intensities. However, testing
for equality, $\lambda^{P}(\cdot)=\lambda^{Q}(\cdot)$, is not flexible
enough. For example, when studying the spatial variation in disease
risk, the diseased population is only a small fraction compared to
the control population; naturally, the corresponding observed patterns
will differ significantly in the overall counts of points---yet this
is irrelevant to the substantive question. The more appropriate null
hypothesis posits that there exists a constant $c$ such that $\lambda^{P}(\cdot)=c\lambda^{Q}(\cdot)$.
Equality within a constant factor means that the intensities have
the same\emph{ functional form} of spatial variation. To avoid dealing
with the nuisance parameter $c$, one can normalize the intensities
to integrate to 1, giving rise to probability distributions $p(\cdot)$
and $q(\cdot)$ over the region $A$; in \cite{cucala_thesis,FirstOrderPatComp}
these are called \emph{the densities of event locations}. Now, our
null hypothesis is equivalent to the equality $p(\cdot)=q(\cdot)$,
which is an instance of the two-sample hypothesis testing problem
(see, e.g. \cite{ANDERSON199441}).

In practice it is desirable to have nonparametric hypothesis testing
approaches to pattern comparison that: a) capture a particular aspect
of difference; b) can be applied to both single and replicated patterns;
and c) do not depend on resampling methods for (re-)calibration. Early
nonparametric tests for pattern comparison \cite{DigglePatComp1,UteHahnJASA}
probe for differences in the $K$-functions \cite{RipleyKFunction}
of point patterns. Being based on a second-order property, the detected
differences conflate the spatial variation in intensities with the
interaction properties. Concentrating on the first-order properties,
\cite{rel_risk1,rel_risk2,rel_risk3} estimate the logarithm of the
ratio between the intensities using kernel density estimation. Other
approaches rely on count of events \cite{ANDRESEN2009333,ALBAFERNANDEZ2016352}
or normalized count of events \cite{ZHANG201772} within pre-specified
areas. The recent work \cite{FirstOrderPatComp} detects differences
in the first-order structure by looking at the $L^{2}$-distance between
kernel density estimates of the probability distributions above (i.e.
$p(\cdot)$ and $q(\cdot)$). All of these truly first-order comparison
approaches are limited to single patterns, and with the exception
of \cite{ZHANG201772} they are calibrated with resampling methods.
The latter issue can result in prohibitive computation costs in industrial
settings where thousands of pattern comparisons may be needed together
with requiring high precision $p$-values to account for multiple
testing corrections. 

In this paper, we introduce an approach that leverages the kernel
mean embedding (KME) \cite{mmd,MMD_review} to test for the equality
$p(\cdot)=q(\cdot)$, which allows us to detect differences in the
first-order structure of point patterns. Our approach is based on
introducing an approximate version of the kernel mean embedding, aKME.
While the original KME is infinite-dimensional and implicit, our approximate
kernel mean embedding is finite-dimensional and comes with explicit
closed-form formulas. With the help of aKME, we reduce the pattern
comparison problem to the comparison of means in the Euclidean space.

\begin{figure}
\begin{centering}
\includegraphics[width=0.85\textwidth]{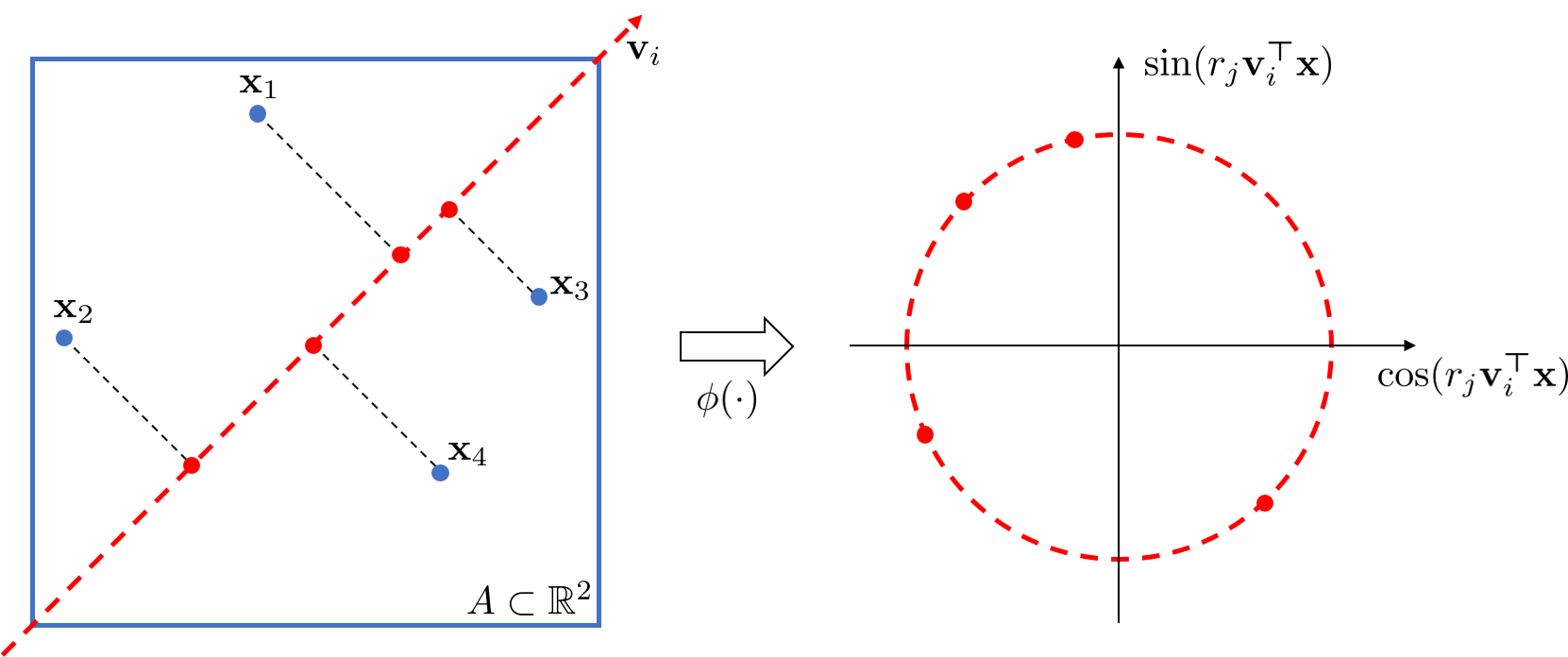}
\par\end{centering}
\caption{\label{fig:rff_pic}Our approximate mean kernel embedding can be seen
as projecting the point pattern onto a line followed by the application
of $\sin/\cos$ functions with the appropriate frequencies. The overall
dimensionality of the resulting embedding is $D=2\times\mathrm{number\,of\,projections}\times\mathrm{number\,of\,frequencies}$.
Figure adapted from \cite{Rahimi:2007:RFL:2981562.2981710}.}
\end{figure}

The resulting pattern comparison test is surprisingly simple and a
complete implementation is provided in the appendix. The computation
of aKME is illustrated in Figure \ref{fig:rff_pic}. First, the points
in the pattern are projected onto a line, which is followed by the
application of $\sin/\cos$ functions with a specific frequency; this
step can be seen as wrapping the line onto a circle of some radius.
The resulting $\sin/\cos$ values are separately averaged to give
two numbers that provide a ``fingerprint'' of the point pattern
behavior with respect to the direction of the line and the scale that
corresponds to the frequency (i.e. circle circumference). The process
is repeated with a multitude of lines and frequencies; assuming $m$
lines and $\ell$ frequencies per line, we obtain $m\ell$ such fingerprints;
these are concatenated together to give an overall $D=2m\ell$ dimensional
aKME. Finally, to compare patterns, we compare their aKMEs by applying
$t$-tests on each coordinate of the embedding. We combine the resulting
$D$ $p$-values into a single overall $p$-value using one of the
recently introduced $p$-value combination techniques, such as harmonic
mean \cite{HarmonicP1958,HarmonicP} or Cauchy combination test \cite{CauchyP},
leading to well-calibrated and powerful tests as confirmed by the
simulations. 

The connection to the original KME guides the choice of the parameters
for this construction and provides approximation guarantees that are
crucial to the consistency of the hypothesis testing. The main advantages
of the proposed approach are that it can be applied to both single
and replicated pattern comparisons, and that neither bootstrap nor
permutation procedures are needed to obtain or calibrate the $p$-values.
In addition, being based on $t$-tests, one can compute Bayes factors
for each of the involved tests allowing to quantify evidence supporting
the hypothesis of difference for each directionality/scale represented
in aKME; one can also report the averaged Bayes factor as an overall
summary of this evidence. 

The ideas developed in this paper are in line with the recent surge
of interest in applying reproducing kernel Hilbert space techniques
to the comparison of probability distributions. For example, the Maximum
Mean Discrepancy (MMD) is a measure of divergence between distributions
\cite{mmd} which has already found numerous applications in statistics
and machine learning. This has led to the general approach of kernel
mean embeddings, see for example the recent review \cite{MMD_review}
and citations therein. Some of these notions can be traced back and
seen as closely related to N-distances \cite{klebanov92} and energy
distances \cite{BARINGHAUS2004190,SZEKELY200558}. Our approximate
embedding has its roots in the Random Fourier Features \cite{Rahimi:2007:RFL:2981562.2981710},
its improvements \cite{KernelApprox1,KernelApprox2,SkolkovoKernelApprox},
and its application to the MMD \cite{Fastmmd}; the scheme we propose
in this paper is tailored to the two-dimensional setting, and has
the ability to provide higher-order approximations. There has already
been some interest in applying the reproducing kernel methodology
to spatial point processes, the roots going back to the 1980s \cite{bartoszynski1981,silverman1982}
and more recently in \cite{flaxman2017,jitkrittum_linear-time_2017,SteinPapangelou}.
We discuss some of the connections between reproducing kernel machinery
and kernel density estimation based methods commonly used with spatial
point patterns in Section \ref{sec:Review-of-Kernel}.

The main contributions of this paper are the proposed approximate
kernel mean embedding (Section \ref{sec:Approximate-Kernel-Mean})
and the hypothesis testing framework for comparison of point patterns
(Section \ref{sec:Hypothesis-Tests}). After investigating the empirical
properties of the resulting tests on simulated data (Section \ref{subsec:Simulations}),
we present applications of the methodology to two real world datasets
(Section \ref{subsec:Real-World-Data}).

\section{\label{sec:Review-of-Kernel}Preliminaries}

\paragraph*{Kernel Mean Embedding}

Mathematically rigorous development of the kernel mean embedding requires
the machinery of the reproducing kernel Hilbert spaces, and the interested
reader is referred to \cite{MMD_review}. For our purposes, it will
be sufficient to concentrate on the dual version of the definition
where the kernel mean embedding is expressed in terms of the feature
maps.

Given a data instance $\mathbf{x}\in\mathcal{X}$ (in our context
$\mathbf{x}$ will be a point in some region of $\mathbb{R}^{2}$),
a nonlinear transformation $\phi:\mathcal{X}\rightarrow\mathcal{F}$
can be used to lift this point into a feature space $\mathcal{F}$
that is usually high/infinite dimensional. When the entire dataset
is lifted like this, the hope is that a good feature map would reveal
structures in the dataset that were not apparent in the original space
$\mathcal{X}$. For example, in the context of classification in machine
learning, two classes that are hard to separate in the original space
may turn out linearly separable in the feature space. This is the
foundation of kernel based approaches in machine learning (see e.g.
\cite{ElementsStatisticalLearning}), with the idea going back to
\cite{cover_theorem}. 

Assuming that an inner product $\langle\cdot,\cdot\rangle$ is available
on $\mathcal{F}$, with each feature map $\phi$ one can associate
the kernel function $k:\mathcal{X}\times\mathcal{X}\rightarrow\mathbb{R}$
defined by $k(\mathbf{x},\mathbf{y})=\langle\phi(\mathbf{x}),\phi(\mathbf{y})\rangle$.
It turns out that the converse is true as well: for each positive
semi-definite (psd) kernel function $k(\cdot,\cdot)$ there exist
a corresponding inner-product space $\mathcal{F}$ together with a
feature map $\phi$ satisfying $k(\mathbf{x},\mathbf{y})=\langle\phi(\mathbf{x}),\phi(\mathbf{y})\rangle$.

For a given psd kernel $k(\cdot,\cdot)$, the kernel mean embedding
(KME) of a probability distribution $p$ over $\mathcal{X}$ is defined
as the expectation 
\[
\mu^{p}=\mathbb{E}_{\mathbf{x}\sim p}[\phi(\mathbf{x})]\in\mathcal{F}.
\]
An important property of this embedding is that when $k(\cdot,\cdot)$
is a \emph{characteristic} kernel, the embedding is injective \cite{MMD_review}.
In other words, for two distributions $p$ and $q$ over $\mathcal{X}$,
the equality $\mu^{p}=\mu^{q}$ is satisfied if and only if $p=q$.
This means that in the context of two-sample hypothesis testing, intuitively,
the problem of testing whether $p=q$ can be reduced to testing $\mu^{p}=\mu^{q}$,
albeit in the infinite-dimensional space $\mathcal{F}$. 

\paragraph*{Two-Sample Tests with KME}

The absence of an explicit formula for $\phi$ and its infinite-dimensionality
pose obstacles to devising an actual hypothesis testing procedure;
fortunately, these can be overcome by using the kernel trick. Consider
the distance (as induced by the inner-product in $\mathcal{F}$) between
KMEs of $p$ and $q$, which is known as the Maximum Mean Discrepancy
(MMD) \cite{mmd}:
\begin{equation}
\mathrm{MMD}^{2}(p,q)=\Vert\mu^{p}-\mu^{q}\Vert^{2}.\label{eq:mmd}
\end{equation}
The kernel trick, namely the application of the equality $k(\mathbf{x},\mathbf{y})=\langle\phi(\mathbf{x}),\phi(\mathbf{y})\rangle$,
allows to express the MMD in terms of $k(\cdot,\cdot)$ without requiring
an explicit formula for the feature map $\phi$, namely, 
\begin{equation}
\mathrm{MMD}^{2}(p,q)=\E_{\mathbf{x},\mathbf{x}'\sim p}[k(\mathbf{x},\mathbf{x}')]-2\E_{\mathbf{x}\sim p,\boldsymbol{\mathbf{y}}\sim q}[k(\mathbf{x},\boldsymbol{\mathbf{y}})]+\E_{\mathbf{y},\mathbf{y}'\sim q}[k(\mathbf{y},\mathbf{y}')].\label{eq:mmd2}
\end{equation}

In practice this statistic is estimated from the samples by replacing
the expectations with the sample means, with a nuance that the first
and third terms can either include the diagonal terms or exclude them;
the former gives a biased and the latter gives an unbiased estimator.
Large enough values of $\mathrm{MMD}^{2}(p,q)$ provide evidence against
the null hypothesis of $p=q$. 

This has been an extremely fruitful approach to two-sample testing
as witnessed by the vast amount of the follow-up work, see the references
in \cite{mmd,MMD_review}. The distribution of MMD under the null
hypothesis is not asymptotically normal nor has a practically useful
explicit form except in big data settings where one can compute an
incomplete estimator of MMD that is asymptotically normal; these cases
are described in generality in \cite{MMDyamada2018post}. As a result,
in typical scenarios some kind of approximation or resampling procedure
is required to perform the test, see \cite{mmd} and pointers therein. 

\paragraph*{Connections to Kernel Density Estimation}

The above described machinery has a close connection to kernel density
estimation commonly used with spatial point patterns. For example,
let us consider the two-sample test statistic proposed in \cite{ANDERSON199441},
which was used by \cite{Duong8382} in a three-dimensional setting
and later adapted to the spatial point pattern comparison problem
in \cite{FirstOrderPatComp}. This statistic is constructed by first
obtaining kernel density estimates of the two distributions and then
computing the squared $L^{2}$-distance between the density estimates.
Its connection to MMD is described in \cite[Section 3.3.1]{mmd} and
can be exemplified as follows: if, say, the kernel density estimate
is based on the Gaussian kernel of bandwidth $\sigma$, then this
statistic is equivalent to the MMD computed with the same kernel but
of bandwidth $\sqrt{2}\sigma$. Overall, MMD statistic is strictly
more general than the statistic of \cite{ANDERSON199441} because
it allows using a larger variety of kernels; another benefit is the
stronger consistency guarantees that stem from the kernel mean embedding
interpretation. Similarly to the MMD, the statistic of \cite{ANDERSON199441}
is not asymptotically normal under the null; nevertheless, a normal
approximation was suggested in \cite{Duong8382}. However, this approximation
leads to severe calibration problems: \cite{FirstOrderPatComp} reported
conservativeness of the resulting test, which led them to rely on
bootstrap for calibration. 

Of course, instead of the $L^{2}$-distance, other distances can be
computed between the kernel density estimates to obtain statistics
that capture complementary aspects of the differences between distributions.
Similarly, one can realize that the MMD captures only one aspect of
the discrepancy between the kernel mean embeddings. For example, the
kernel Fisher discriminant analysis (KFDA) test statistic \cite{KernelFisherDA1}
is another statistic that can be computed via the kernel trick and
used to test the equality $\mu^{p}=\mu^{q}$. Similarly to the relationship
between MMD and the squared $L^{2}$-distance, the KFDA is connected
to the $\chi^{2}$-distance between kernel density estimates. An insightful
review of this type of connections can be found in \cite{KernelFisherDA2}. 

\section{\label{sec:Approximate-Kernel-Mean}Approximate Kernel Mean Embedding}

Instead of relying on the kernel trick, in this section we take an
orthogonal path to avoiding the infinite-dimensionality of the kernel
mean embedding. Namely, specializing to the case $\mathcal{X}=\mathbb{R}^{2}$,
we construct a finite-dimensional \emph{approximate} feature map $\phi:\mathbb{R}^{2}\rightarrow\mathbb{R}^{D}$
such that $k(\mathbf{x},\mathbf{y})\approx\phi(\mathbf{x})^{\boldsymbol{\top}}\phi(\mathbf{y})$.
As a result, testing $p=q$ can be reduced to testing $\mu^{p}=\mu^{q}$
in the $D$-dimensional Euclidean space. 

Since it allows obtaining closed form formulas, for the rest of the
paper we will concentrate on the Gaussian kernel $k(\mathbf{x},\mathbf{y})=e^{-\Vert\mathbf{x}-\mathbf{y}\Vert^{2}/2\sigma^{2}}$
for $\mathbf{x},\mathbf{y}\in\mathbb{R}^{2}$. To avoid clutter in
the derivations the kernel bandwidth $\sigma$ is assumed to be $1$;
the case of general $\sigma$ is mentioned below. Our goal is to construct
an approximate feature map for the Gaussian kernel tailored to the
two-dimensional case. As a starting point, following the notation
from \cite{SkolkovoKernelApprox} let us write the Gaussian kernel
as

\[
k(\mathbf{x},\mathbf{y})=\frac{1}{2\pi}\int_{-\infty}^{\infty}\int_{-\infty}^{\infty}e^{-\frac{\mathbf{w}^{\boldsymbol{\top}}\mathbf{w}}{2}}\cos(\mathbf{w}^{\boldsymbol{\top}}(\mathbf{x}-\mathbf{y}))d\mathbf{w}=\frac{1}{2\pi}\int_{-\infty}^{\infty}\int_{-\infty}^{\infty}e^{-\frac{\mathbf{w}^{\boldsymbol{\top}}\mathbf{w}}{2}}\eta(\mathbf{w}^{\boldsymbol{\top}}\mathbf{x})^{\boldsymbol{\top}}\eta(\mathbf{w}^{\boldsymbol{\top}}\mathbf{y})d\mathbf{w},
\]
where $\eta(\cdot)=\left[\begin{array}{cc}
\cos(\cdot) & \quad\sin(\cdot)\end{array}\right]{}^{\boldsymbol{\top}}$. To compress the notation, let $f(\mathbf{w})=\eta(\mathbf{w}^{\boldsymbol{\top}}\mathbf{x})^{\boldsymbol{\top}}\eta(\mathbf{w}^{\boldsymbol{\top}}\mathbf{y})$
where the dependence on $\mathbf{x}$ and $\mathbf{y}$ is suppressed
from the notation. We change to the radial coordinates, and notice
that $f$ is an even function to obtain:

\[
k(\mathbf{x},\mathbf{y})=\frac{1}{2\pi}\int_{0}^{2\pi}\int_{0}^{\infty}re^{-\frac{r^{2}}{2}}f(r\mathbf{z_{\theta}})drd\mathbf{\theta}=\frac{1}{\pi}\int_{0}^{\pi}\int_{0}^{\infty}re^{-\frac{r^{2}}{2}}f(r\mathbf{z_{\theta}})drd\mathbf{\theta},
\]
where $\mathbf{z}_{\theta}$ is a unit vector in $\mathbb{R}^{2}$
making an angle of $\theta$ with the $x$-axis. We approximate the
integration with respect to $\theta$ by the simple mean over a set
of equidistant values $\theta_{i}=(i-1)\pi/m$, $i=1,2,....,m$. Let
$\mathbf{v}_{i}=\mathbf{z}_{\theta_{i}}$ to get 
\[
k(\mathbf{x},\mathbf{y})\approx\sum_{i=1}^{m}\int_{0}^{\infty}re^{-\frac{r^{2}}{2}}f(r\mathbf{v}_{i})dr.
\]

Now we use a radial version of Gauss-Hermite quadrature to compute
the integral over $r$ \cite{jackel-2005a}. For a given integer $\ell\geq1$,
denoting the quadrature roots by $\{r_{j}\}_{j=1}^{\ell}$ and the
corresponding weights by $\{\omega_{j}\}_{j=1}^{\ell}$, we obtain\footnote{The values of the roots and weights can be retrieved from the following
URL:

http://www.jaeckel.org/RootsAndWeightsForRadialGaussHermiteQuadrature.cpp}
\[
k(\mathbf{x},\mathbf{y})\approx\sum_{i=1}^{m}\sum_{j=1}^{\ell}\omega_{j}f(r_{j}\mathbf{v}_{i})=\sum_{i=1}^{m}\sum_{j=1}^{\ell}\omega_{j}\eta(r_{j}\mathbf{v}_{i}^{\boldsymbol{\top}}\mathbf{x})^{\boldsymbol{\top}}\eta(r_{j}\mathbf{v}_{i}^{\boldsymbol{\top}}\mathbf{y}).
\]
The crucial point is that the terms containing $\mathbf{x}$ and $\mathbf{y}$
are untangled, which allows us to write $k(\mathbf{x},\mathbf{y})$
approximately as a dot product in the embedding space: $k(\mathbf{x},\mathbf{y})\approx\phi_{m,\ell}(\mathbf{x})^{\top}\phi_{m,\ell}(\mathbf{y})$,
where $\phi_{m,\ell}:\mathbb{R}^{2}\rightarrow\mathbb{R}^{D}$, $D=2m\ell$
is explicitly given by 
\[
\phi_{m,\ell}(\mathbf{x})=[...,\sqrt{\omega_{j}}\sin(r_{j}\mathbf{v}_{i}^{\boldsymbol{\top}}\mathbf{x}),\sqrt{\omega_{j}}\cos(r_{j}\mathbf{v}_{i}^{\boldsymbol{\top}}\mathbf{x}),...]^{\top},
\]
where in total there are $m\ell$ sine/cosine pairs corresponding
to all combinations of $i$ and $j$. We note that the case of the
Gaussian kernel with the width $\sigma$ can be accommodated by replacing
$\mathbf{x}$ with $\mathbf{x}/\sigma$. To avoid committing to a
single $\sigma$, in practice, one can construct multiple feature
maps corresponding to different values of the kernel width, and concatenate
the resulting vectors together to obtain a higher dimensional embedding
(e.g. using three different width values results in a $3D$-dimensional
embedding). 

The connection to Random Fourier Features \cite{Rahimi:2007:RFL:2981562.2981710}
should be obvious from the form of our embedding. Note that the construction
in \cite{Rahimi:2007:RFL:2981562.2981710} and its improvements \cite{KernelApprox1,KernelApprox2,SkolkovoKernelApprox}
are geared towards high-dimensional data and, thus, depend on stochasticity
to avoid the curse of dimensionality. In contrast, our construction
is tailored to the two-dimensional setting and is based on a deterministic
quadrature rule that quickly provides a high-order approximation.
For example, considering a region $A\subset\mathbb{R}^{2}$ the approximation
error of the kernel function can be bounded as 
\[
\vert k(\mathbf{x},\mathbf{y})-\phi_{m,\ell}(\mathbf{x})^{\top}\phi_{m,\ell}(\mathbf{y})\vert\leq C_{1}\frac{\Vert x-y\Vert}{m}+C_{2}\gamma_{\ell}\frac{\Vert x-y\Vert^{2\ell}}{(2\ell)!}\leq\epsilon(m,\ell,\mathrm{diam}(A)),
\]
for all $\mathbf{x},\mathbf{y}\in A$. The first term comes from the
classical Riemann sum approximation error and the second term is the
standard form for general Gaussian quadrature (Theorem 3.6.24 in \cite{quadraturebook});
the powers of $\Vert x-y\Vert$ arise from differentiation of the
integrand. When using kernel width of $\sigma$, the error depends
on $\mathrm{diam}(A)/\sigma$. Our bound reflects the quality of the
quadrature rules involved in the approximation scheme, and so is conceptually
different from the probabilistic bound in \cite{Rahimi:2007:RFL:2981562.2981710}.
There is quantitative difference as well: their dimension $D$ to
achieve error $\epsilon$ scales roughly as $\epsilon^{-2}\log\epsilon$,
whereas ours as $\epsilon^{-1}\log\epsilon$. As a result, the proposed
embedding is able to capture more information about the distribution
for the same number of dimensions, resulting in more powerful tests.
Nevertheless, the hypothesis tests proposed in Section \ref{sec:Hypothesis-Tests}
do generalize to Random Fourier Feature based embeddings and, consequently,
can be applied in high-dimensional settings. 

Having introduced the approximate feature map, we can define the \emph{approximate}
kernel mean embedding (aKME) of a probability distribution $p(\cdot)$
over the region $A\subset\mathbb{R}^{2}$ as $\mathrm{aKME}_{m,\ell}(p)=\mu_{m,\ell}^{p}=\mathbb{E}_{\mathbf{x}\sim p}[\phi_{m,\ell}(\mathbf{x})]\in\mathbb{R}^{D}$.
For two distributions $p$ and $q$ over $A$,
\begin{equation}
\mathrm{MMD}^{2}(p,q)\approx\Vert\mathrm{aKME}_{m,\ell}(p)-\mathrm{aKME}_{m,\ell}(q)\Vert^{2},\label{eq:mmd_akme}
\end{equation}
where in contrast to Eq. (\ref{eq:mmd}) the embeddings on the right-hand
side are computed via aKME. The approximation error can be estimated
by replacing each term of Eq. (\ref{eq:mmd2}) with their approximations
($\E_{\mathbf{x},\mathbf{x}'\sim p}[k(\mathbf{x},\mathbf{x'})]\approx\E_{\mathbf{x},\mathbf{x}'\sim p}[\phi_{m,\ell}(\mathbf{x})^{\top}\phi_{m,\ell}(\mathbf{x}')]$
and so on), which incurs an overall error upper-bounded by $4\epsilon(m,\ell,\mathrm{diam}(A))$.
The expectations decouple ($\E[\phi_{m,\ell}(\mathbf{x})^{\top}\phi_{m,\ell}(\mathbf{x}')]=\E_{x\sim p}[\phi_{m,\ell}(\mathbf{x})^{\top}]\E_{x'\sim p}[\phi_{m,\ell}(\mathbf{x}')]$
and so on) yielding Eq. (\ref{eq:mmd_akme}) after some algebra; see
\cite{Fastmmd} for a similar argument in the Random Fourier Features
context. 

For a large class of probability distributions, $p\neq q$ implies
$\mathrm{MMD}^{2}(p,q)>0$ \cite{mmd}. Thus, by taking $m$ and $\ell$
large enough so that the approximation error $4\epsilon(m,\ell,\mathrm{diam}(A))$
of Eq. (\ref{eq:mmd_akme}) is small enough we will obtain $\Vert\mathrm{aKME}_{m,\ell}(p)-\mathrm{aKME}_{m,\ell}(q)\Vert^{2}>0$
which implies $\mathrm{aKME}_{m,\ell}(p)\neq\mathrm{aKME}_{m,\ell}(q)$.
In other words, if two distributions are different, this difference
will be discerned by the aKME of high enough dimensionality. This
property of our construction will be crucial when proving the consistency
of the proposed hypothesis tests. 

\paragraph*{Remark}

To simplify notation, in the subsequent sections we will drop the
subscripts $m$ and $\ell$ from all of the approximate constructs.

\section{\label{subsec:Spatial-Point-Pattern-aKME}Spatial Point Pattern aKME}

The goal of this section is to introduce the aKME for a point process
and show that it can be estimated in an unbiased manner from the realizations
of the point process. As a preliminary, we have to compare the notions
of the first-order intensity and the density of event locations for
spatial point processes. While for an inhomogeneous Poisson process
these two are equivalent up to a normalization, in general there are
differences that should be taken into consideration when conducting
replicated pattern comparisons. 

\paragraph*{Intensity versus Location Density}

Let $P$ be a point process on the region $A\subset\mathbb{R}^{2}$.
The first-order intensity function $\lambda^{P}(\cdot$) of this process
is defined as

\[
\lambda^{P}(\mathbf{x})=\lim_{|dx|\to0}\frac{\E[N(d\mathbf{x})]}{\vert d\mathbf{x}\vert},
\]
where $d\mathbf{x}$ is a neighborhood around $\mathbf{x}$, and $N(\cdot)$
is the count of events and $\vert\cdot\vert$ is the area. The definition
of the density of event locations, or simply, location density function
$p(\cdot)$, is as follows: 

\[
p(\mathbf{x})=\lim_{|d\mathbf{x}|\to0}\frac{\E[N(d\mathbf{x})/N(A)]}{\vert d\mathbf{x}\vert},
\]
note the normalization by the overall number of points. It is easy
to see that $p(x)$ integrates to 1 over $A$, and as such is a probability
density function on $A$. 

To clarify the connection between $\lambda^{P}$ and $p$ we can rewrite
the definition by conditioning on the number of events in a point
pattern:

\begin{equation}
p(\mathbf{x})=\lim_{|d\mathbf{x}|\to0}\frac{\E[N(d\mathbf{x})/N(A)]}{\vert d\mathbf{x}\vert}=\E_{n}\left[\frac{1}{n}\lim_{|d\mathbf{x}|\to0}\frac{\E_{N(A)=n}[N(d\mathbf{x})]}{\vert d\mathbf{x}\vert}\right]=\E_{n}\left[\frac{1}{n}\lambda^{P}(\mathbf{x}|N(A)=n)\right],\label{eq:intensity_density}
\end{equation}
where $\lambda^{P}(\mathbf{x}|N(A)=n)$ is the first-order intensity
conditioned on the total number of points in the pattern being equal
to $n$.

For an inhomogeneous Poisson process, the intensity and the location
density are equivalent up to normalization. Indeed, consider a Poisson
process $P$ on $A$ with intensity $\lambda^{P}(\cdot$) and denote
$\beta=\int_{A}\lambda^{P}(\mathbf{x})d\mathbf{x}$. Conditioning
a Poisson process to have $n$ points gives the binomial process with
intensity $n\lambda^{P}(\cdot)/\beta$. Plugging this into Eq. (\ref{eq:intensity_density}),
we get $p(\cdot)=\lambda^{P}(\cdot)/\beta$.

Turning to the non-Poisson case, we see that if the condition
\begin{equation}
\lambda^{P}(\mathbf{x}|N(A)=n)=n\frac{\lambda^{P}(\mathbf{x})}{\int_{A}\lambda^{P}(\mathbf{x})d\mathbf{x}}\quad\forall n\:\mathrm{such\:that}\:\mathbf{\mathrm{Prob}}(N(A)=n)>0\label{eq:condition}
\end{equation}
holds, the above argument can be repeated verbatim to give the equivalence
between intensity and location density: $p(\cdot)=\lambda^{P}(\cdot)/\int_{A}\lambda^{P}(\mathbf{x})d\mathbf{x}$.
We will provide a detailed discussion of this requirement at the end
of this section.

\paragraph*{Spatial Point Pattern aKME}

The aKME of the point process $P$ is defined via the embedding of
its location density function $p(\cdot)$, namely, $\mu^{P}=\mu^{p}=\mathbb{E}_{\mathbf{x}\sim p}[\phi(\mathbf{x})]\in\mathbb{R}^{D}.$
In practice, we only have access to a point process via its realizations,
and so, we must be able to estimate the aKME in an unbiased manner
from a point pattern; this is the focus of the following discussion.

Consider a realization $\boldsymbol{X}=\{\mathbf{x}_{1},\mathbf{x}_{2},...,\mathbf{x}_{n}\}$
of the point process $P$. The estimator that we will be interested
in is the one that replaces the expectation appearing in the aKME
formula by the sample mean:
\begin{equation}
\hat{\mu}^{\boldsymbol{X}}=\frac{1}{n}\sum_{j=1}^{n}\phi(\mathbf{x}_{j})\in\mathbb{R}^{D}.\label{eq:SampleMeanEst}
\end{equation}
We require this estimate to be unbiased, $\mathbb{E}_{\mathbf{X}\sim P}[\hat{\mu}^{\boldsymbol{X}}]=\mu^{p}$,
where the expectation is taken over the realizations of the process
$P$. Unbiasedness is essential as it ensures that the estimate $\hat{\mu}^{\boldsymbol{X}}$
captures information about the location density function rather than
some other properties of the point process.

Next, we prove that unbiasedness holds for any point process that
satisfies the condition of Eq. (\ref{eq:condition}). We can write,
\[
\mathbb{E}_{\boldsymbol{X}\sim P}[\hat{\mu}^{\boldsymbol{X}}]=\mathbb{E}_{n}\left[\frac{1}{n}\mathbb{E}_{\boldsymbol{X}\sim P,|\mathbf{X}|=n}\left[\sum_{j=1}^{n}\phi(\mathbf{x}_{j})\right]\right].
\]
Using Campbell's formula and Eq. (\ref{eq:condition}), we get 
\[
\frac{1}{n}\mathbb{E}_{\boldsymbol{X}\sim P,|\mathbf{X}|=n}\left[\sum_{j=1}^{n}\phi(\mathbf{x}_{j})\right]=\frac{1}{n}\int_{A}\phi(\mathbf{x})\lambda^{P}(\mathbf{x}|N(A)=n)d\mathbf{x}=\int_{A}\phi(\mathbf{x})p(\mathbf{x})d\mathbf{x}.
\]
Substituting this back, 
\[
\mathbb{E}_{\boldsymbol{X}\sim P}[\hat{\mu}^{\boldsymbol{X}}]=\mathbb{E}_{n}\left[\int_{A}\phi(\mathbf{x})p(\mathbf{x})d\mathbf{x}\right]=\int_{A}\phi(\mathbf{x})p(\mathbf{x})d\boldsymbol{\mathbf{x}}=\mathbb{E}_{\mathbf{x}\sim p}[\phi(\mathbf{x})]=\mu^{p},
\]
which proves that the estimator is unbiased.

\paragraph*{Discussion}

The equivalence between the first-order intensity and the location
density functions implied by Eq. (\ref{eq:condition}) is important
for our hypothesis tests, as we replace the comparison of intensities
by the comparison densities; we also saw that Eq. (\ref{eq:condition})
is at the core of estimating aKME in an unbiased manner. Thus, the
validity of our hypothesis testing framework in Section \ref{sec:Hypothesis-Tests}
hinges on this condition, and so it requires further discussion. Our
single pattern comparison test assumes that patterns are sampled from
(in-)homogeneous Poisson processes. We already saw that inhomogeneous
Poisson processes satisfy this property, which validates our single
comparison test. On the other hand, our replicated pattern comparison
test can be used for non-Poisson processes, and so one has to ascertain
that the condition of Eq. (\ref{eq:condition}) holds. 

Here we focus on a practical discussion and leave theoretical analysis
of this requirement to the future work. We loosely express the requirement
of Eq. (\ref{eq:condition}) as follows:\emph{ conditioning on the
number of points in the pattern does not have an effect on the functional
form of the first-order intensity} (i.e. it induces a change by a
constant factor). The simplicity of this requirement makes it possible
to use domain knowledge to assess how likely it is to hold in practice.
For example, non-Poisson point processes are common in plant biology
applications; the functional form of the first-order intensity is
determined by the environmental factors such as light conditions,
level of water underground, and quality of soil. Thus, conditioning
on the total number of plants in the region of interest should not
affect the first-order intensity except for an overall constant factor,
which makes it plausible to assume that the condition highlighted
above holds. 

A similar argument can be made for patterns of a certain type of crime
in a town on a fixed day of the week. However, if we were to consider
the patterns without regard for the day of the week then violations
of Eq. (\ref{eq:condition}) are possible. To this end, consider a
scenario where: 1) the number of crimes on weekends is typically much
lower than on weekdays, and 2) crimes on weekdays typically happen
in the residential areas and on weekends in the business areas. Thus,
conditioning on the number of events in a pattern effectively splits
the patterns by weekday/weekend, and the corresponding event intensities
concentrate on different parts of the town, resulting in the failure
of the requirement above. 

In data rich regimes one can use the following strategy to asses the
requirement of Eq. (\ref{eq:condition}). One can group the collection
of patterns by the number of events, so that patterns in the same
group do not differ drastically in the number of events. Next, our
replicated pattern comparison test can be applied to each pair of
groups to determine if there are inter-group differences in the location
densities. If no significant differences are found, then one can assume
that the overall set of patterns satisfies the condition of Eq. (\ref{eq:condition}).
The approximate validity of this approach stems from the following
observation. An obvious case where the requirement trivially holds
is when every realization of the process has the same number of events.
Based on this, we conjecture that in practical situations having small
differences in the counts of events in the patterns should not lead
to big violations of the requirement allowing us to use the test.

\section{\label{sec:Hypothesis-Tests}Comparing Spatial Point Patterns with
aKME }

Consider two point processes $P$ and $Q$ in $A\subset\mathbb{R}^{2}$
with the first-order intensity functions given by $\lambda^{P}(\cdot$)
and $\lambda^{Q}(\cdot)$. We would like to test the null hypothesis
of whether there exists a constant $c$ such that $\lambda^{P}(\cdot)=c\lambda^{Q}(\cdot)$.
Equality up to a constant factor means that the intensities of the
two processes have the same functional form. This is different from
testing $\lambda^{P}(\cdot)=\lambda^{Q}(\cdot)$ because our null
hypothesis can hold true even if the realizations from $P$ and $Q$
have vastly differing numbers of events. We start with the Poisson
case that allows testing based on a single realization per process
and proceed to replicated pattern comparison that is applicable more
generally.

\paragraph*{Single Pattern Comparison}

Assume that we observe two patterns $\boldsymbol{X}=\{\mathbf{x}_{1},\mathbf{x}_{2},...,\mathbf{x}_{n_{1}}\}$
and $\boldsymbol{Y}=\{\mathbf{y}_{1},\mathbf{y}_{2},...,\boldsymbol{\mathbf{y}}_{n_{2}}\}$
from two inhomogeneous Poisson processes $P$ and $Q$ on the region
$A\subset\mathbb{R}^{2}$ with the true intensity functions given
by $\lambda^{P}(\cdot$) and $\lambda^{Q}(\cdot)$. We would like
to test the null hypothesis of whether there exists a constant $c$
such that $\lambda^{P}(\cdot)=c\lambda^{Q}(\cdot)$. Using the equivalence
between intensities and location densities of Poisson processes proved
in Section \ref{subsec:Spatial-Point-Pattern-aKME}, the location
density functions $p(\cdot)$ and $q(\cdot)$ are obtained from the
intensities $\lambda^{P}(\cdot$) and $\lambda^{Q}(\cdot)$ via normalization
(e.g. $p(\cdot)=\lambda^{P}(\cdot)/\int_{A}\lambda^{P}(\mathbf{x})d\mathbf{x}$).
Clearly, the null hypothesis above can be written as $p=q$.

Applying the aKME machinery, we reduce the test to checking whether
$\mu^{p}=\mu^{q}$ in the $D$-dimensional Euclidean space. While
the classic Hotelling's $T^{2}$ test seems like a natural choice,
yet given the high-dimensionality of our embedding this becomes very
data hungry; in practice the computation of the precision matrix turns
out to be unstable. We also tried using a test geared to the high-dimensional
setting  \cite{chen2010}, but it failed to control the size of the
test due to the nonlinear relationships between the coordinates of
the approximate feature map which breaks the multivariate normality
assumption. 

To avoid these issues we perform the test by applying $t$-tests independently
to each coordinate and combining the resulting $p$-values as explained
below. More precisely, for each coordinate $i=1,2,...,D$, by Poisson
assumption we can consider $\{\phi_{i}(\mathbf{x}_{j}),\mathbf{x}_{j}\in\boldsymbol{X}\}$
and $\{\phi_{i}(\mathbf{y}_{j}),\mathbf{y}_{j}\in\boldsymbol{Y}\}$
to be two independent samples of sizes $n_{1}$ and $n_{2}$ respectively;
here $\phi_{i}$ is the $i$-th coordinate of the approximate feature
map. Our goal is to compare the sample means $\hat{\mu}_{i}^{\boldsymbol{X}}$
and $\hat{\mu}_{i}^{\boldsymbol{Y}}$. Note that since the approximate
feature maps are expressed in terms of sines and cosines, their range
is bounded, and so the classical Central Limit Theorem can be applied
to deduce the approximate normality of the sample means $\hat{\mu}_{i}^{\boldsymbol{X}}$
and $\hat{\mu}_{i}^{\boldsymbol{Y}}$ \emph{when the sample sizes
are big enough}. As a result, we can apply the Behrens-Fisher-Welch
$t$-test (without assuming equality of variances) to obtain the corresponding
$p$-value $p_{i}$. 

All together we end up with a set of $p$-values $p_{1},p_{2},...,p_{D}$
one per coordinate of the aKME. In principle, one can resort to multiple
testing procedures to reject or retain the global null. However, it
is useful to compute an overall $p$-value for the the global null
hypothesis $\mu^{p}=\mu^{q}$ so that it can be used in downstream
multiple testing procedures. To obtain such an overall $p$-value
we use one of the $p$-value combination approaches, see Eqs. (\ref{eq:HMP})
and (\ref{eq:CauchyP}) at the end of this section. \textit{Using
the approximation properties of aKME we prove in Appendix \ref{sec:consistency}
that the proposed test is consistent.}

\paragraph*{Replicated Pattern Comparison}

In the replicated pattern setting we can abandon the Poisson assumption;
the only assumption on the involved point processes is that condition
of Eq. (\ref{eq:condition}) holds. Here, for two such general processes
$P$ and $Q$ on the region $A\subset\mathbb{R}^{2}$ with true intensity
functions given by $\lambda^{P}(\cdot$) and $\lambda^{Q}(\cdot)$,
we observe a set of patterns $P^{*}=\{\boldsymbol{X}_{1},\boldsymbol{X}_{2},...,\boldsymbol{X}_{m_{1}}\}$
and $Q^{*}=\{\boldsymbol{Y}_{1},\boldsymbol{Y}_{2},...,\boldsymbol{Y}_{m_{2}}\}$
independently drawn from $P$ and $Q$ respectively. Our goal is to
test whether $\lambda^{P}(\cdot)=c\lambda^{Q}(\cdot)$. As proved
in Section \ref{subsec:Spatial-Point-Pattern-aKME}, the equivalence
between intensities and location densities holds, and so the problem
can be reduced to testing $p=q$ exactly as before.

In this setting, we can consider the collection of each coordinate
$i$ of the aKME embeddings $\{\hat{\mu}_{i}^{\boldsymbol{X}_{k}},\boldsymbol{X}_{k}\in P^{*}\}$
and $\{\hat{\mu}_{i}^{\boldsymbol{Y}_{k}},\boldsymbol{Y}_{k}\in Q^{*}\}$
as independent samples of sizes $m_{1}$ and $m_{2}$. Being the means
of sines and cosines, the range of each coordinate is bounded, but
in contrast to the previous case the means coming from different replications
are not identically distributed due to the varying number of events
in each replication. Uniform boundedness and independence allow us
to apply the Lyapunov/Lindeberg version of the Central Limit Theorem
\cite[Chapter 27]{CLT_Ref} to deduce the approximate normality of
the sample means \emph{when the sample sizes (i.e. the number of replications
of patterns) are big enough}. This allows us to test for the equality
of the $i$-th coordinates $\mu_{i}^{p}=\mu_{i}^{q}$ using the $t$-test.
Here, the validity of the $t$-test stems from independence of replications
and holds for general processes; this should be contrasted to the
single pattern comparison test above which derives its validity from
the independence of samples within a single pattern in the Poisson
process. The set of $D$ $p$-values obtained from the $t$-tests
are combined using one of the $p$-value combination approaches discussed
below to obtain an overall $p$-value. 

We finish with an important remark about the nature of this test.
The point processes $P$ and $Q$ do not have to be of the same type
for the null hypothesis to hold. The test is concerned only with the
functional form of the intensity, and not the type of the process.
For example, if $P$ is an inhibition process and $Q$ is a cluster
process, then the null hypothesis still holds as long as the first-order
intensity functions of these processes are the same up to a constant
factor; having it otherwise would have implied that the test conflates
higher order properties with the first order properties.

\paragraph*{Combining P-Values}

To obtain an overall $p$-value for the tests described above, we
need to combine the per-coordinate $p$-values $p_{1},p_{2},...,p_{D}$
in a manner that is robust to the dependencies between them. We found
experimentally that classical $p$-value combination approaches such
as Fisher's and Stouffer's methods (see, e.g. \cite{pvalue_combo1,pvalue_combo2})
fail to give well-calibrated $p$-values likely due to their strong
reliance on the independence assumption. In contrast, the recently
introduced combination techniques, harmonic mean $p$-value \cite{HarmonicP1958,HarmonicP}
and Cauchy combination test \cite{CauchyP} resulted in well-calibrated
tests with better power than simple alternatives such as the Bonferroni
adjustment for multiple testing. We quickly review these approaches
and provide some insight into their behavior and the relationship
between them. 

The harmonic mean $p$-value combination approach defines an overall
$p$-value by
\begin{equation}
p^{H}=H\left(\frac{D}{\frac{1}{p_{1}}+\frac{1}{p_{2}}+\cdots+\frac{1}{p_{D}}}\right),\label{eq:HMP}
\end{equation}
where $H(\cdot)$ is a function whose precise form is described in
\cite{HarmonicP}. Since $H(x)\approx x$ for small values of $x$,
$p^{H}$ is approximately the harmonic mean of $p_{1},p_{2},...,p_{D}$.
On the other hand, the Cauchy combination test defines an overall
$p$-value by the formula

\begin{equation}
p^{C}=\frac{1}{\pi}\cot^{-1}\left(\frac{\cot\pi p_{1}+\cot\pi p_{2}+\cdots+\cot\pi p_{D}}{D}\right),\label{eq:CauchyP}
\end{equation}
where $\cot$ is the cotangent function, $\cot x=\tan(\pi/2-x)$.
In contrast to classical combination techniques, both of these approaches
are shown to be robust when there are dependencies between the individual
$p$-values $p_{1},p_{2},...,p_{D}$ \cite{HarmonicP,CauchyP}. 

Curiously, these two methods behave very similarly for small $p$-values.
Indeed, we can rewrite Eq. (\ref{eq:CauchyP}) as $D\cot\pi p^{C}=(\cot\pi p_{1}+\cot\pi p_{2}+\cdots+\cot\pi p_{D})$.
For small $x$, the approximation $\cot x\approx1/x$ can be used,
and canceling out $\pi$s on both sides we get $D/p^{C}\approx1/p_{1}+1/p_{2}+\cdots+1/p_{D}$.
It follows that $p^{C}$ is approximately the harmonic mean of $p_{1},p_{2},...,p_{D}$,
and, therefore, $p^{C}\approx p^{H}$. 

\begin{figure}
\begin{centering}
\includegraphics[width=1\textwidth]{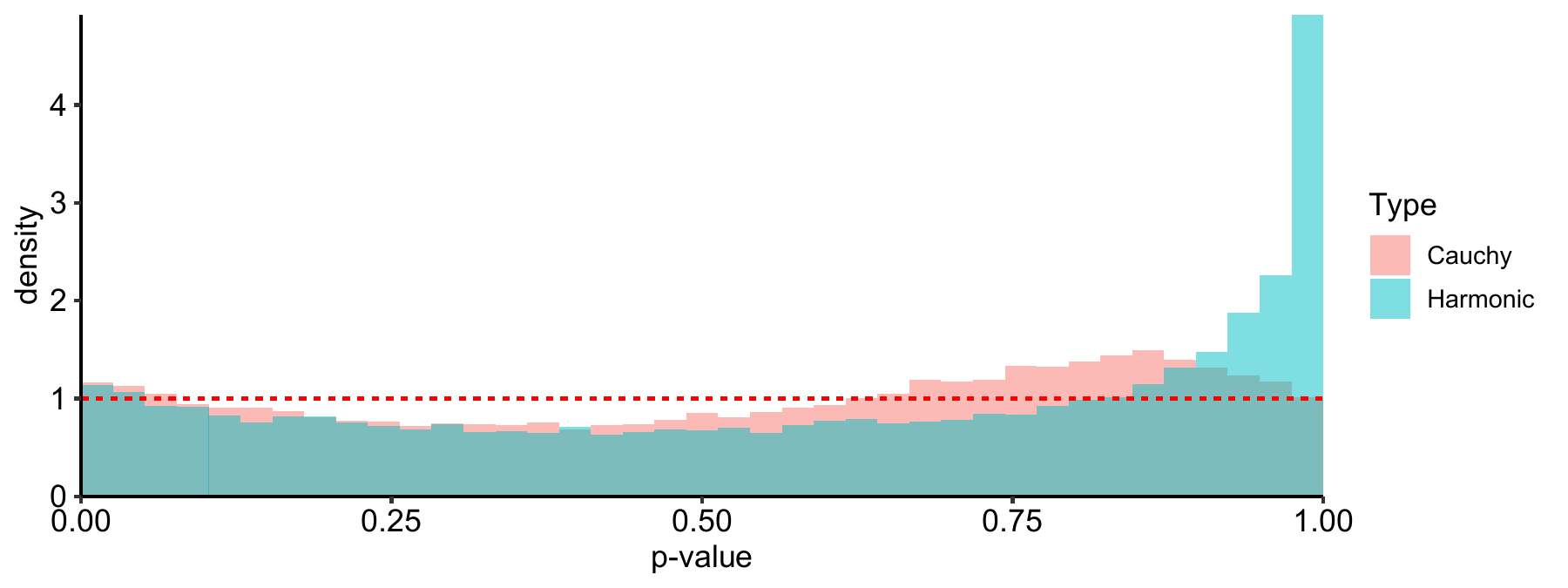}
\par\end{centering}
\caption{\label{fig:cauchy_vs_harmonic}Distribution of combined $p$-values
under the null for the Poisson process experiment in Section \ref{subsec:Simulations}.}
\end{figure}

To gain some more intuition about the behavior of these techniques
we can look at the distribution of the combined $p$-values under
the null. Figure \ref{fig:cauchy_vs_harmonic} depicts the histogram
of combined $p$-values corresponding to the Poisson process experiment
in Section \ref{subsec:Simulations} when the null holds. It can be
seen that the methods behave very similarly for small values of $p$
as expected from the approximation argument above. Both of the combination
approaches result in unit density for small values of $p$, which
makes them equally suitable for hypothesis testing (c.f. ``Size''
portion of Table \ref{tab:Homog-vs.-inhomog}). Interestingly, the
harmonic mean approach leads to overcrowding of $p$-values near $1$,
whereas the Cauchy combination $p$-values stay more or less uniformly
distributed. The large $p$-value behavior can in general be ignored
in the context of hypothesis testing. However, it can be a consideration
when using adaptive false discovery control techniques based on the
\emph{mirroring technique} \cite{barber2015,arias-castro2017} that
relies on symmetric distribution of $p$-values near $0$ and $1$.
Otherwise, the harmonic mean approach has a number of advantages,
including a Bayesian model-averaging interpretation and its being
an inherently multilevel test \cite{HarmonicP}. 

Due to their similar behavior in the relevant regime, we use the harmonic
mean $p$-value approach in all of the experiments presented in the
paper. The sample source code provided in Appendix \ref{sec:R-Implementation}
demonstrates the calculation for both of the $p$-value combination
techniques.

\paragraph*{Adjusting $p$-values}

While the $p$-value combination methods above give excellent results
in practice and are fast to compute, their approximate nature and
the resulting non-uniformity of $p$-values over the entire range
can potentially be a problem in some applications. Here we present
two adjustment approaches to obtaining guaranteed size control. Let
us emphasize that we do not use any adjustments in the experiments
reported in the main text.

The first approach is based on the worst-case performance of $p$-value
combination techniques \cite{Vovk_combo} and will lead to rather
conservative tests. For example, using the Bonferroni adjustment is
equivalent to using the combined $p$-value $p^{B}=D\min\{p_{1},p_{2},...,p_{D}\}$.
For the pure harmonic mean combination (i.e. without the application
of function $H$ in Eq. (\ref{eq:HMP})) it holds that $a^{H}(D)p^{H}$
provides guaranteed size control for some factor $a^{H}(D)$ that
depends on the number of the $p$-values being combined \cite{Vovk_combo};
there is no closed-form formula for this factor, but it satisfies
$a^{H}(D)\leq e\ln D$ and $\lim_{D\to\infty}a^{H}(D)/\ln D=1$.

Another method is to adjust the $p$-values using resampling. Focusing
on the harmonic approach, our test statistic would be the harmonic
combination $p$-value. We generate $B$ samples of the harmonic $p$-value
$\{p_{b}^{H}\}_{b=1}^{B}$ from the null distribution. Given the observed
value of $p^{H}$ for the comparison at hand, we compute the adjusted
harmonic $p$-value via the formula $p^{AH}=(\vert\{b:p_{b}^{H}\leq p^{H}\}\vert+1)/(B+1)$.
The generation of the null distribution is guided by the test and
a choice between randomized permutation or bootstrap. For single pattern
comparison, we can either randomly shuffle points across the two point
patterns or sample points with replacement from the aggregation of
the two patterns (smoothed bootstrap can be used by estimating the
location density of the aggregate pattern); for replicated pattern
comparison we can similarly either shuffle the patterns across the
groups or sample patterns with replacement from the overall pattern
pool. Note that the resampling approach can be used in cases where
the Poisson assumption (required for single pattern comparison) fails
as long as one can generate a valid null distribution by techniques
such as tiled resampling \cite{HALL1985231}.

\paragraph*{Bayes Factors}

\begin{figure}
\begin{centering}
\includegraphics[width=1\textwidth]{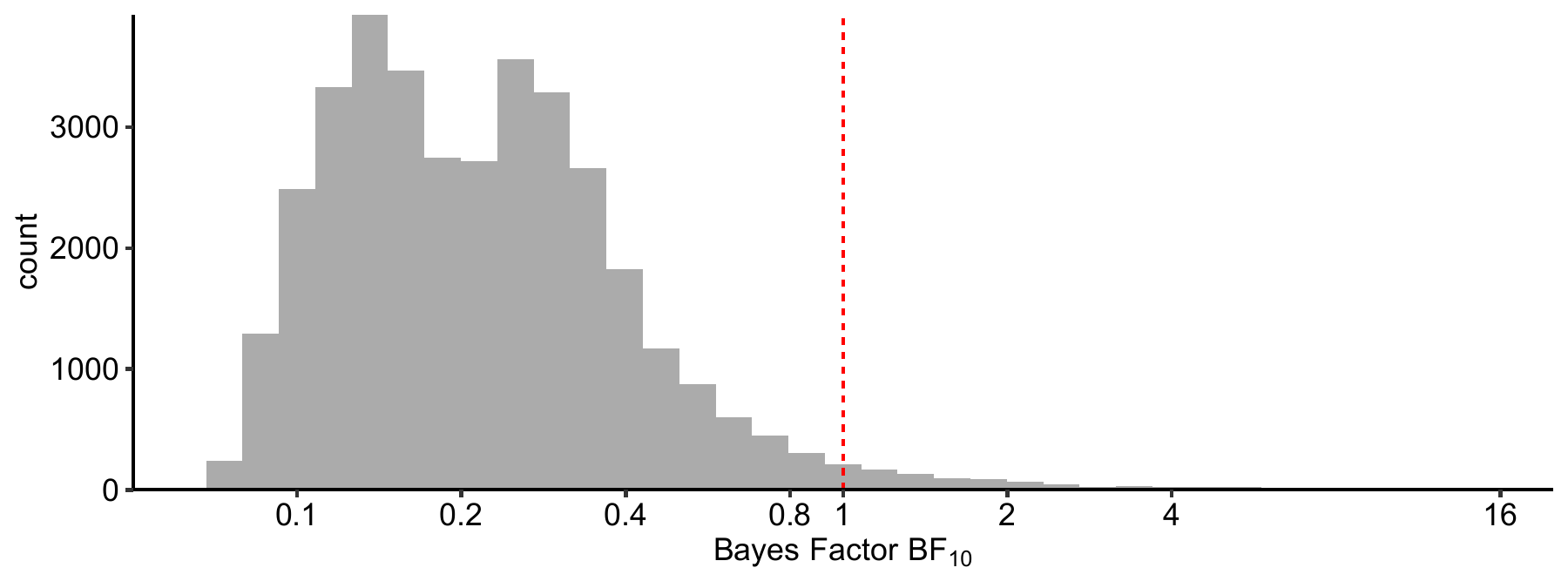}
\par\end{centering}
\caption{\label{fig:bf_null}Distribution of mean Bayes factors under the null
for the Poisson process experiment in Section \ref{subsec:Simulations}.
The $x$-axis has logarithmic scale. As expected, these values are
overwhelmingly below 1, shown by a red line.}
\end{figure}

Each dimension of the aKME captures the behavior of the point pattern
in a specific direction at a specific scale. Thus, Bayesian approaches
can potentially be used for hypothesis testing with priors that capture
the beliefs for differences along each dimension of the aKME. To illustrate
this idea, we follow a straightforward approach and rely on the existing
default priors for $t$-test (e.g. \cite{ttestBF}) to compute Bayes
factors \cite{Jeffreys61,doi:10.1080/01621459.1995.10476572} for
each dimension of the aKME. Next, we combine these Bayes factors following
the work of \cite{vovk2019combining} via arithmetic mean to obtain
what Vovk and Wang call an $e$-value; we will refer to this quantity
as the mean Bayes factor and will denote it by $\overline{\mathrm{BF}}_{10}$.
In essence, $\overline{\mathrm{BF}}_{10}$ captures the overall evidence
against the null hypothesis and under the null it is expected to be
smaller than 1. We confirm this by computing the mean Bayes factors
for the null cases of Poisson process experiment in Section \ref{subsec:Simulations}.
Figure \ref{fig:bf_null} shows the histogram of the resulting $\overline{\mathrm{BF}}_{10}$
values and demonstrates that under the null they overwhelmingly take
values smaller than 1. We will report mean Bayes factors for the real
world experiments in Section \ref{subsec:Real-World-Data}.

\section{Experiments}

Our goal in this section is to investigate the size and the power
of the proposed tests. We also demonstrate two applications to real
world data. The aKME embedding is constructed using four radial projections
and four roots for the polar Gauss-Hermite formula (i.e. $m=4$, $\ell=4$
in the notation of Section \ref{sec:Approximate-Kernel-Mean}) resulting
in $D=32$. To avoid the selection of the kernel width parameter,
we concatenate together aKMEs corresponding to $\sigma=1/16,1/8,$
and $1/4$ when the point pattern domain is the unit square; the dimensionality
of the concatenated aKME is $3\times32=96$. To motivate these choices
of $\sigma$, notice that the spatial scales represented in aKME correspond
to the periods of the involved trigonometric functions and have the
form $2\pi\sigma/r_{j}$. Consulting the values for $r_{j}$ from
the source code in Appendix \ref{sec:R-Implementation}, we see that
the corresponding spatial scales vary from approximately $0.1$ to
$4$, which we expect would reveal differences between patterns considered
here. All of these values except for two are under $\sqrt{2}$ (diagonal
of the square); while those two largest values may seem redundant,
they are required for revealing the gross differences between patterns
when they exist.

\subsection{\label{subsec:Simulations}Simulations}

\paragraph*{Single Pattern Comparison}

We generate two types of inhomogeneous Poisson processes (Linear and
Sine) on the unit square $[0,1]\times[0,1]$ with the intensity functions
given by $\exp(-\gamma x)$ and $\exp(-\gamma\sin(2\pi x))$, where
$x$ is the abscissa; the form of these is borrowed from \cite{YongtaoGuan2008}.
Here, $\gamma$ takes the values of $\gamma=1,2,3$; the intensity
functions are normalized so that the expected number of points per
realization is $100$, $400$ or $800$. For each pair of parameter
settings being compared the model is simulated 2000 times.

\newlength{\mytabcolsep}
\setlength{\mytabcolsep}{\tabcolsep}
\setlength{\tabcolsep}{1pt}
\renewcommand{\arraystretch}{0.9}

\begin{table}
\begin{centering}
\begin{tabular}{lccccccccccccccccccccc}
\toprule 
 &  &  &  &  &  &  & \multicolumn{3}{c}{{\footnotesize{}This Paper}} &  &  &  & \multicolumn{3}{c}{{\footnotesize{}Zhang and Zhuang}} &  &  &  & \multicolumn{3}{c}{{\footnotesize{}Fuentes-Santos et al.}}\tabularnewline
\midrule 
{\footnotesize{}Model} & {\footnotesize{}$\beta_{1}$} & {\footnotesize{}$\beta_{2}$} & {\footnotesize{}$\lambda$} &  &  &  & {\footnotesize{}$\alpha=0.01$} & {\footnotesize{}$\alpha=0.05$} & {\footnotesize{}$\alpha=0.1$} &  &  &  & {\footnotesize{}$\alpha=0.01$} & {\footnotesize{}$\alpha=0.05$} & {\footnotesize{}$\alpha=0.1$} &  &  &  & {\footnotesize{}$\alpha=0.01$} & {\footnotesize{}$\alpha=0.05$} & {\footnotesize{}$\alpha=0.1$}\tabularnewline
\midrule
\midrule 
\textbf{\footnotesize{}Size} &  &  &  &  &  &  &  &  &  &  &  &  &  &  &  &  &  &  &  &  & \tabularnewline
{\footnotesize{}Linear} & {\footnotesize{}1} & {\footnotesize{}1} & {\footnotesize{}100} &  &  &  & {\footnotesize{}0.008} & {\footnotesize{}0.058} & {\footnotesize{}0.108} &  &  &  & {\footnotesize{}0.007} & {\footnotesize{}0.035} & {\footnotesize{}0.068} &  &  &  & {\footnotesize{}0.014} & {\footnotesize{}0.062} & {\footnotesize{}0.111}\tabularnewline
 &  & {\footnotesize{}1} & {\footnotesize{}400} &  &  &  & {\footnotesize{}0.009} & {\footnotesize{}0.045} & {\footnotesize{}0.098} &  &  &  & {\footnotesize{}0.010} & {\footnotesize{}0.045} & {\footnotesize{}0.088} &  &  &  & {\footnotesize{}0.014} & {\footnotesize{}0.052} & {\footnotesize{}0.098}\tabularnewline
 &  & {\footnotesize{}1} & {\footnotesize{}800} &  &  &  & {\footnotesize{}0.008} & {\footnotesize{}0.050} & {\footnotesize{}0.104} &  &  &  & {\footnotesize{}0.010} & {\footnotesize{}0.048} & {\footnotesize{}0.090} &  &  &  & {\footnotesize{}0.010} & {\footnotesize{}0.051} & {\footnotesize{}0.091}\tabularnewline
 & {\footnotesize{}2} & {\footnotesize{}2} & {\footnotesize{}100} &  &  &  & {\footnotesize{}0.011} & {\footnotesize{}0.054} & {\footnotesize{}0.106} &  &  &  & {\footnotesize{}0.007} & {\footnotesize{}0.038} & {\footnotesize{}0.076} &  &  &  & {\footnotesize{}0.007} & {\footnotesize{}0.053} & {\footnotesize{}0.094}\tabularnewline
 &  & {\footnotesize{}2} & {\footnotesize{}400} &  &  &  & {\footnotesize{}0.008} & {\footnotesize{}0.052} & {\footnotesize{}0.099} &  &  &  & {\footnotesize{}0.010} & {\footnotesize{}0.048} & {\footnotesize{}0.091} &  &  &  & {\footnotesize{}0.012} & {\footnotesize{}0.053} & {\footnotesize{}0.102}\tabularnewline
 &  & {\footnotesize{}2} & {\footnotesize{}800} &  &  &  & {\footnotesize{}0.010} & {\footnotesize{}0.044} & {\footnotesize{}0.090} &  &  &  & {\footnotesize{}0.011} & {\footnotesize{}0.048} & {\footnotesize{}0.088} &  &  &  & {\footnotesize{}0.006} & {\footnotesize{}0.056} & {\footnotesize{}0.100}\tabularnewline
 & {\footnotesize{}3} & {\footnotesize{}3} & {\footnotesize{}100} &  &  &  & {\footnotesize{}0.010} & {\footnotesize{}0.042} & {\footnotesize{}0.090} &  &  &  & {\footnotesize{}0.006} & {\footnotesize{}0.030} & {\footnotesize{}0.060} &  &  &  & {\footnotesize{}0.009} & {\footnotesize{}0.046} & {\footnotesize{}0.095}\tabularnewline
 &  & {\footnotesize{}3} & {\footnotesize{}400} &  &  &  & {\footnotesize{}0.012} & {\footnotesize{}0.058} & {\footnotesize{}0.106} &  &  &  & {\footnotesize{}0.010} & {\footnotesize{}0.038} & {\footnotesize{}0.084} &  &  &  & {\footnotesize{}0.011} & {\footnotesize{}0.050} & {\footnotesize{}0.100}\tabularnewline
 &  & {\footnotesize{}3} & {\footnotesize{}800} &  &  &  & {\footnotesize{}0.012} & {\footnotesize{}0.048} & {\footnotesize{}0.088} &  &  &  & {\footnotesize{}0.010} & {\footnotesize{}0.050} & {\footnotesize{}0.092} &  &  &  & {\footnotesize{}0.010} & {\footnotesize{}0.058} & {\footnotesize{}0.112}\tabularnewline
{\footnotesize{}Sine} & {\footnotesize{}1} & {\footnotesize{}1} & {\footnotesize{}100} &  &  &  & {\footnotesize{}0.010} & {\footnotesize{}0.054} & {\footnotesize{}0.100} &  &  &  & {\footnotesize{}0.006} & {\footnotesize{}0.031} & {\footnotesize{}0.076} &  &  &  & {\footnotesize{}0.012} & {\footnotesize{}0.048} & {\footnotesize{}0.106}\tabularnewline
 &  & {\footnotesize{}1} & {\footnotesize{}400} &  &  &  & {\footnotesize{}0.014} & {\footnotesize{}0.047} & {\footnotesize{}0.102} &  &  &  & {\footnotesize{}0.009} & {\footnotesize{}0.046} & {\footnotesize{}0.082} &  &  &  & {\footnotesize{}0.014} & {\footnotesize{}0.046} & {\footnotesize{}0.100}\tabularnewline
 &  & {\footnotesize{}1} & {\footnotesize{}800} &  &  &  & {\footnotesize{}0.012} & {\footnotesize{}0.059} & {\footnotesize{}0.106} &  &  &  & {\footnotesize{}0.006} & {\footnotesize{}0.038} & {\footnotesize{}0.076} &  &  &  & {\footnotesize{}0.018} & {\footnotesize{}0.062} & {\footnotesize{}0.119}\tabularnewline
 & {\footnotesize{}2} & {\footnotesize{}2} & {\footnotesize{}100} &  &  &  & {\footnotesize{}0.013} & {\footnotesize{}0.057} & {\footnotesize{}0.096} &  &  &  & {\footnotesize{}0.004} & {\footnotesize{}0.027} & {\footnotesize{}0.068} &  &  &  & {\footnotesize{}0.012} & {\footnotesize{}0.043} & {\footnotesize{}0.088}\tabularnewline
 &  & {\footnotesize{}2} & {\footnotesize{}400} &  &  &  & {\footnotesize{}0.014} & {\footnotesize{}0.058} & {\footnotesize{}0.094} &  &  &  & {\footnotesize{}0.008} & {\footnotesize{}0.034} & {\footnotesize{}0.074} &  &  &  & {\footnotesize{}0.012} & {\footnotesize{}0.048} & {\footnotesize{}0.098}\tabularnewline
 &  & {\footnotesize{}2} & {\footnotesize{}800} &  &  &  & {\footnotesize{}0.008} & {\footnotesize{}0.046} & {\footnotesize{}0.095} &  &  &  & {\footnotesize{}0.006} & {\footnotesize{}0.038} & {\footnotesize{}0.084} &  &  &  & {\footnotesize{}0.011} & {\footnotesize{}0.043} & {\footnotesize{}0.096}\tabularnewline
 & {\footnotesize{}3} & {\footnotesize{}3} & {\footnotesize{}100} &  &  &  & {\footnotesize{}0.008} & {\footnotesize{}0.044} & {\footnotesize{}0.092} &  &  &  & {\footnotesize{}0.008} & {\footnotesize{}0.031} & {\footnotesize{}0.065} &  &  &  & {\footnotesize{}0.012} & {\footnotesize{}0.044} & {\footnotesize{}0.104}\tabularnewline
 &  & {\footnotesize{}3} & {\footnotesize{}400} &  &  &  & {\footnotesize{}0.008} & {\footnotesize{}0.051} & {\footnotesize{}0.090} &  &  &  & {\footnotesize{}0.006} & {\footnotesize{}0.036} & {\footnotesize{}0.077} &  &  &  & {\footnotesize{}0.012} & {\footnotesize{}0.058} & {\footnotesize{}0.120}\tabularnewline
 &  & {\footnotesize{}3} & {\footnotesize{}800} &  &  &  & {\footnotesize{}0.010} & {\footnotesize{}0.046} & {\footnotesize{}0.091} &  &  &  & {\footnotesize{}0.010} & {\footnotesize{}0.052} & {\footnotesize{}0.100} &  &  &  & {\footnotesize{}0.008} & {\footnotesize{}0.059} & {\footnotesize{}0.126}\tabularnewline
\midrule 
\textbf{\footnotesize{}Power} &  &  &  &  &  &  &  &  &  &  &  &  &  &  &  &  &  &  &  &  & \tabularnewline
{\footnotesize{}Linear} & {\footnotesize{}1} & {\footnotesize{}2} & {\footnotesize{}100} &  &  &  & {\footnotesize{}0.085} & {\footnotesize{}0.206} & {\footnotesize{}0.310} &  &  &  & {\footnotesize{}0.086} & {\footnotesize{}0.232} & {\footnotesize{}0.336} &  &  &  & {\footnotesize{}0.056} & {\footnotesize{}0.182} & {\footnotesize{}0.278}\tabularnewline
 &  & {\footnotesize{}2} & {\footnotesize{}400} &  &  &  & {\footnotesize{}0.644} & {\footnotesize{}0.816} & {\footnotesize{}0.882} &  &  &  & {\footnotesize{}0.613} & {\footnotesize{}0.820} & {\footnotesize{}0.893} &  &  &  & {\footnotesize{}0.332} & {\footnotesize{}0.578} & {\footnotesize{}0.704}\tabularnewline
 &  & {\footnotesize{}2} & {\footnotesize{}800} &  &  &  & {\footnotesize{}0.972} & {\footnotesize{}0.992} & {\footnotesize{}0.996} &  &  &  & {\footnotesize{}0.958} & {\footnotesize{}0.994} & {\footnotesize{}0.996} &  &  &  & {\footnotesize{}0.728} & {\footnotesize{}0.862} & {\footnotesize{}0.921}\tabularnewline
 &  & {\footnotesize{}3} & {\footnotesize{}100} &  &  &  & {\footnotesize{}0.587} & {\footnotesize{}0.762} & {\footnotesize{}0.848} &  &  &  & {\footnotesize{}0.524} & {\footnotesize{}0.752} & {\footnotesize{}0.844} &  &  &  & {\footnotesize{}0.401} & {\footnotesize{}0.606} & {\footnotesize{}0.739}\tabularnewline
 &  & {\footnotesize{}3} & {\footnotesize{}400} &  &  &  & {\footnotesize{}1.000} & {\footnotesize{}1.000} & {\footnotesize{}1.000} &  &  &  & {\footnotesize{}1.000} & {\footnotesize{}1.000} & {\footnotesize{}1.000} &  &  &  & {\footnotesize{}0.982} & {\footnotesize{}0.996} & {\footnotesize{}0.998}\tabularnewline
 &  & {\footnotesize{}3} & {\footnotesize{}800} &  &  &  & {\footnotesize{}1.000} & {\footnotesize{}1.000} & {\footnotesize{}1.000} &  &  &  & {\footnotesize{}1.000} & {\footnotesize{}1.000} & {\footnotesize{}1.000} &  &  &  & {\footnotesize{}1.000} & {\footnotesize{}1.000} & {\footnotesize{}1.000}\tabularnewline
 & {\footnotesize{}2} & {\footnotesize{}3} & {\footnotesize{}100} &  &  &  & {\footnotesize{}0.070} & {\footnotesize{}0.179} & {\footnotesize{}0.274} &  &  &  & {\footnotesize{}0.060} & {\footnotesize{}0.182} & {\footnotesize{}0.281} &  &  &  & \textcolor{brown}{\footnotesize{}0.040} & \textcolor{brown}{\footnotesize{}0.130} & \textcolor{brown}{\footnotesize{}0.216}\tabularnewline
 &  & {\footnotesize{}3} & {\footnotesize{}400} &  &  &  & {\footnotesize{}0.520} & {\footnotesize{}0.725} & {\footnotesize{}0.808} &  &  &  & {\footnotesize{}0.478} & {\footnotesize{}0.720} & {\footnotesize{}0.819} &  &  &  & \textcolor{brown}{\footnotesize{}0.207} & \textcolor{brown}{\footnotesize{}0.446} & \textcolor{brown}{\footnotesize{}0.587}\tabularnewline
 &  & {\footnotesize{}3} & {\footnotesize{}800} &  &  &  & {\footnotesize{}0.915} & {\footnotesize{}0.968} & {\footnotesize{}0.982} &  &  &  & {\footnotesize{}0.876} & {\footnotesize{}0.958} & {\footnotesize{}0.982} &  &  &  & \textcolor{brown}{\footnotesize{}0.436} & \textcolor{brown}{\footnotesize{}0.716} & \textcolor{brown}{\footnotesize{}0.822}\tabularnewline
{\footnotesize{}Sine} & {\footnotesize{}1} & {\footnotesize{}2} & {\footnotesize{}100} &  &  &  & {\footnotesize{}0.352} & {\footnotesize{}0.608} & {\footnotesize{}0.720} &  &  &  & {\footnotesize{}0.146} & {\footnotesize{}0.336} & {\footnotesize{}0.472} &  &  &  & {\footnotesize{}0.196} & {\footnotesize{}0.388} & {\footnotesize{}0.544}\tabularnewline
 &  & {\footnotesize{}2} & {\footnotesize{}400} &  &  &  & {\footnotesize{}0.997} & {\footnotesize{}0.999} & {\footnotesize{}1.000} &  &  &  & {\footnotesize{}0.962} & {\footnotesize{}0.994} & {\footnotesize{}0.997} &  &  &  & {\footnotesize{}0.842} & {\footnotesize{}0.956} & {\footnotesize{}0.978}\tabularnewline
 &  & {\footnotesize{}2} & {\footnotesize{}800} &  &  &  & {\footnotesize{}1.000} & {\footnotesize{}1.000} & {\footnotesize{}1.000} &  &  &  & {\footnotesize{}1.000} & {\footnotesize{}1.000} & {\footnotesize{}1.000} &  &  &  & {\footnotesize{}0.994} & {\footnotesize{}1.000} & {\footnotesize{}1.000}\tabularnewline
 &  & {\footnotesize{}3} & {\footnotesize{}100} &  &  &  & {\footnotesize{}0.942} & {\footnotesize{}0.985} & {\footnotesize{}0.990} &  &  &  & {\footnotesize{}0.576} & {\footnotesize{}0.806} & {\footnotesize{}0.885} &  &  &  & {\footnotesize{}0.752} & {\footnotesize{}0.934} & {\footnotesize{}0.962}\tabularnewline
 &  & {\footnotesize{}3} & {\footnotesize{}400} &  &  &  & {\footnotesize{}1.000} & {\footnotesize{}1.000} & {\footnotesize{}1.000} &  &  &  & {\footnotesize{}1.000} & {\footnotesize{}1.000} & {\footnotesize{}1.000} &  &  &  & {\footnotesize{}1.000} & {\footnotesize{}1.000} & {\footnotesize{}1.000}\tabularnewline
 &  & {\footnotesize{}3} & {\footnotesize{}800} &  &  &  & {\footnotesize{}1.000} & {\footnotesize{}1.000} & {\footnotesize{}1.000} &  &  &  & {\footnotesize{}1.000} & {\footnotesize{}1.000} & {\footnotesize{}1.000} &  &  &  & {\footnotesize{}1.000} & {\footnotesize{}1.000} & {\footnotesize{}1.000}\tabularnewline
 & {\footnotesize{}2} & {\footnotesize{}3} & {\footnotesize{}100} &  &  &  & {\footnotesize{}0.050} & {\footnotesize{}0.190} & {\footnotesize{}0.308} &  &  &  & \textcolor{magenta}{\footnotesize{}0.014} & \textcolor{magenta}{\footnotesize{}0.056} & \textcolor{magenta}{\footnotesize{}0.116} &  &  &  & \textcolor{brown}{\footnotesize{}0.028} & \textcolor{brown}{\footnotesize{}0.133} & \textcolor{brown}{\footnotesize{}0.242}\tabularnewline
 &  & {\footnotesize{}3} & {\footnotesize{}400} &  &  &  & {\footnotesize{}0.768} & {\footnotesize{}0.912} & {\footnotesize{}0.942} &  &  &  & \textcolor{magenta}{\footnotesize{}0.118} & \textcolor{magenta}{\footnotesize{}0.296} & \textcolor{magenta}{\footnotesize{}0.424} &  &  &  & \textcolor{brown}{\footnotesize{}0.174} & \textcolor{brown}{\footnotesize{}0.424} & \textcolor{brown}{\footnotesize{}0.574}\tabularnewline
 &  & {\footnotesize{}3} & {\footnotesize{}800} &  &  &  & {\footnotesize{}0.994} & {\footnotesize{}1.000} & {\footnotesize{}1.000} &  &  &  & \textcolor{magenta}{\footnotesize{}0.511} & \textcolor{magenta}{\footnotesize{}0.766} & \textcolor{magenta}{\footnotesize{}0.854} &  &  &  & \textcolor{brown}{\footnotesize{}0.456} & \textcolor{brown}{\footnotesize{}0.726} & \textcolor{brown}{\footnotesize{}0.846}\tabularnewline
\bottomrule
\end{tabular}
\par\end{centering}
\caption{\label{tab:Homog-vs.-inhomog}Rejection rates for the single pattern
comparison test when the Poisson assumption holds.}
\end{table}

\setlength{\tabcolsep}{\mytabcolsep}
\renewcommand{\arraystretch}{1.0}

Table \ref{tab:Homog-vs.-inhomog} lists under the heading ``This
Paper'' the rejection rates of our test at the nominal levels $\alpha=0.01,0.05,$
and $0.1$. The top part of the table corresponds to the case where
the patterns being compared come from the same intensity model. The
resulting rejection rates give the size of the test; note that all
of them are close to the nominal sizes, which confirms that our test
is well-calibrated if only slightly liberal. In the bottom half of
the table the patterns being compared come from different intensity
models; the corresponding rejection rates give the power of the test. 

We compare our test to the Kolmogorov--Smirnov type test of \cite{ZHANG201772}
and the resampling based test of \cite{FirstOrderPatComp}. Zhang
and Zhuang actually propose two test statistics, $T_{1}$ and $T_{2}$;
in our setting $T_{1}$ performs better in terms of power, so we present
results based on $T_{1}$. Our computations are based on their code
that includes a more efficient estimator of the test statistic for
Poisson processes. Experiments for Fuentes-Santos et al. are based
on their code with a modification that we provide the theoretical
null location density functions (instead of estimating them via kernel
density estimation) in order to save computation time, which potentially
may give the method some advantage.

The last six columns of Table \ref{tab:Homog-vs.-inhomog} show the
rejection rates for Zhang \& Zhuang and Fuentes-Santos et al. tests.
Being based on an asymptotic result, the Zhang \& Zhuang test can
be seen to be conservative for smaller sample sizes. Keeping this
in mind, for most of the pattern comparisons all of the three tests
have similar power. An interesting exception happens when comparing
Sine patterns for the parameter values $\gamma=2$ and $\gamma=3$.
While our test quickly reaches high power with the increasing sample
size, it can be seen that both Zhang \& Zhuang and Fuentes-Santos
et al. tests struggle in this setting (values highlighted in magenta
and brown). Moreover, Fuentes-Santos et al. test has a similar issue
for Linear $\gamma=2$ versus $\gamma=3$ comparison. We believe that
for Zhang \& Zhuang test this may be related to their test statistic
being based on the maximum difference of normalized counts, which
can be overwhelmed by the noise coming from the dense areas, thereby,
ignoring the true differences in the sparse areas of the patterns.
Given the connection of Fuentes-Santos et al. method to the MMD based
testing as described in Section \ref{sec:Review-of-Kernel}, we speculate
that this power difference is an indication that treating the dimensions
of the aKME individually is preferable over testing using Euclidean
distance in the aKME space (which is approximately equal to the MMD).

In the appendix we provide a similar table of comparison between the
proposed test and its adjusted version. In Table \ref{tab:harmonic-vs-adjusted}
the values under ``Adjusted Harmonic'' heading are computed using
the resampling adjustment; the difference in terms of power between
harmonic and adjusted harmonic methods are minimal especially for
large $\lambda$. This shows that the highlighted power difference
in the previous paragraph cannot be attributed to the approximate
nature of the $p$-value combination approach, but is a true feature
of the proposed test.

\begin{table}
\begin{centering}
\begin{tabular}{llccc}
\toprule 
 & Model & $\alpha=0.01$ & $\alpha=0.05$ & $\alpha=0.1$\tabularnewline
\midrule
\midrule 
\textbf{Size} & Hardcore-1 & 0.012 & 0.046 & 0.090\tabularnewline
 & Hardcore-2 & 0.008 & 0.035 & 0.073\tabularnewline
 & Hardcore-3 & 0.002 & 0.009 & 0.020\tabularnewline
 & Cluster-1 & 0.056 & 0.164 & 0.271\tabularnewline
 & Cluster-2 & 0.160 & 0.372 & 0.508\tabularnewline
 & Cluster-3 & 0.413 & 0.686 & 0.780\tabularnewline
\bottomrule
\end{tabular}
\par\end{centering}
\caption{\label{tab:Robustness}Single pattern testing depends on the validity
of the Poisson assumption. When it is violated, the test size does
not match the nominal rate; inhibition and clustering have opposite
effects on the size.}
\end{table}

\begin{table}
\begin{centering}
\begin{tabular}{llccc}
\toprule 
 & Model & $\alpha=0.01$ & $\alpha=0.05$ & $\alpha=0.1$\tabularnewline
\midrule
\midrule 
\textbf{Size} & Cluster-1 & 0.011 & 0.051 & 0.086\tabularnewline
 & Cluster-2 & 0.014 & 0.048 & 0.086\tabularnewline
 & Cluster-3 & 0.015 & 0.050 & 0.078\tabularnewline
\bottomrule
\end{tabular}
\par\end{centering}
\caption{\label{tab:Robustness-Eff}Size of the single pattern comparison test
when the Poisson assumption is violated due to clustering. Using the
\emph{effective sample size} in the $t$-tests for clustering processes
results in sizes that are close to the nominal rate.}
\end{table}

Next we investigate what happens when the Poisson assumption is violated.
To this end we run comparisons between patterns from homogeneous Poisson
process and patterns from homogeneous non-Poisson processes. This
is the null case, because the functional form of intensity is the
same (constant) for all of these cases. The non-Poisson processes
we consider are generated using the spatstat package \cite{spatstat}
and are as follows. We use Matern's Model II inhibition process with
the inhibition distance parameter setting of $r=0.01,0.02,0.04$;
these are referred to as Hardcore-1,2,3. We also generate realizations
from Matern's cluster process with the radius parameter value of 0.1,
and with the mean number of points parameter $\mu=1,2,4$; these are
called Cluster-1,2,3 respectively. The underlying intensity for all
of the processes is set to yield an average of 100 points per pattern.

Table \ref{tab:Robustness} lists the resulting null rejection rates
which correspond to the size of the test. We see that the inhibition
process results in the test becoming more conservative; we conjecture
that this is true in general, and so, when inhibition can be argued
based on the domain knowledge, the rejections obtained via our test
are still valid. In contrast, clustering quickly leads to an anti-conservative
test. The latter is not surprising, as the clustered points cannot
be seen as independently drawn. One way of looking at this is that
the \emph{effective} sample size in a clustered process is lower than
the number of points in the pattern. 

We investigate whether correcting the $t$-tests for the sample size
can bring the test size to the nominal level. We made a back-of-the-envelope
computation of the effective sample size for Matern's cluster process.
The goal here is to determine the number of parents $n_{\mathrm{eff}}$
that ended up generating the $n$ points seen in the pattern. Since
each parent gets $\mathrm{Poisson}(\mu)$ offsprings and empty clusters
get discarded, the average number of points per cluster is $\E_{c\sim\mathrm{Poisson}(\mu),c\geq1}[c]=\mu/(1-\exp(-\mu))$,
and dividing the total count by this average gives $n_{\mathrm{eff}}=n(1-\exp(-\mu))/\mu$.
The resulting sizes are shown in Table \ref{tab:Robustness-Eff},
and match the nominal rates more closely. 

Of course, in practice, one pattern is not enough to deduce whether
the clustering is due to the heterogeneity of the intensity or due
to clustering; thus, some modeling assumptions on the point process
would be required in order to carry out such a correction. For example,
in plant biology applications one may know the typical radius of a
cluster, which can guide some type farthest point sampling to approximately
determine the number of independent/parent events, and using the latter
as the effective sample size. In cases involving data collection from
units, such as users of a geolocation powered application, the count
of distinct units can be used in lieu of the effective sample size.

\paragraph*{Replicated Pattern Comparison}

Since in the replicated pattern test does not require the Poisson
assumption, we generate seven classes of point patterns first homogeneous
and second non-homogeneous, yielding fourteen types of patterns in
total. For homogeneous versions, we generate CSR, Hardcore-1,2,3,
and Cluster-1,2,3. To obtain the non-homogeneous versions of these,
we start with a homogeneous version and independently thin it with
spatially varying retention probability of $\exp(-x)$; this is based
on \cite{10.2307/4541321}. As a result, we obtain inhomogeneous point
patterns where the intensity has the functional form $\exp(-x)$.
All generation processes were manipulated so that the expected number
of points per pattern is roughly $100$; there are 20 patterns per
group. 

The first row in the Table \ref{tab:replic-pattern} lists the rejection
rates corresponding to the cases when the patterns compared have the
same inhomogeneity type without regard for the model (CSR, Hardcore,
Cluster) of the pattern. This corresponds to the null case, and we
can see that the sizes are close to the nominal levels. The second
row lists the rejection rates when the the inhomogeneity types differ,
and so correspond to the power of the test. As noted before, the replicated
pattern test is concerned only with the functional form of the intensity,
and not the type of the process. For example, when comparing an inhibition
process to a cluster process, the null hypothesis still holds as long
as the first-order intensity functions of these processes are the
same up to a constant factor; having it otherwise would have implied
that the test conflates higher order properties with the first order
properties. Nevertheless, it is instructive to split out from the
aforementioned aggregate table the cases where comparisons happen
between the same model types only, such as Hardcore pattern groups
being compared only amongst themselves and so on. We can see from
Table \ref{tab:replic-pattern-1} that for all of these comparsions
we control the size and obtain high power. 

\begin{table}
\begin{centering}
\begin{tabular}{lccc}
\toprule 
 & $\alpha=0.01$ & $\alpha=0.05$ & $\alpha=0.1$\tabularnewline
\midrule
\midrule 
Size & 0.008 & 0.054 & 0.107\tabularnewline
\midrule 
Power & 0.976 & 0.995 & 0.997\tabularnewline
\bottomrule
\end{tabular}
\par\end{centering}
\caption{\label{tab:replic-pattern}Rejection rates for the replicated pattern
comparison test for assorted model type comparisons.}
\end{table}

\begin{table}[ht]
\begin{centering}
\begin{tabular}{llccc}
\toprule 
 & Model & $\alpha=0.01$ & $\alpha=0.05$ & $\alpha=0.1$\tabularnewline
\midrule
\midrule 
Size & Poisson & 0.003 & 0.048 & 0.101\tabularnewline
 & Hardcore & 0.007 & 0.056 & 0.101\tabularnewline
 & Cluster & 0.010 & 0.049 & 0.088\tabularnewline
\midrule 
Power & Poisson & 1.000 & 1.000 & 1.000\tabularnewline
 & Hardcore & 1.000 & 1.000 & 1.000\tabularnewline
 & Cluster & 0.938 & 0.982 & 0.991\tabularnewline
\bottomrule
\end{tabular}
\par\end{centering}
\caption{\label{tab:replic-pattern-1}Rejection rates for the replicated pattern
comparison test for the same model type comparisons.}
\end{table}

\subsection{\label{subsec:Real-World-Data}Real-World Data}

\paragraph*{Single Pattern: Cancer Data}

\begin{figure}
\begin{centering}
\includegraphics[width=0.55\textwidth]{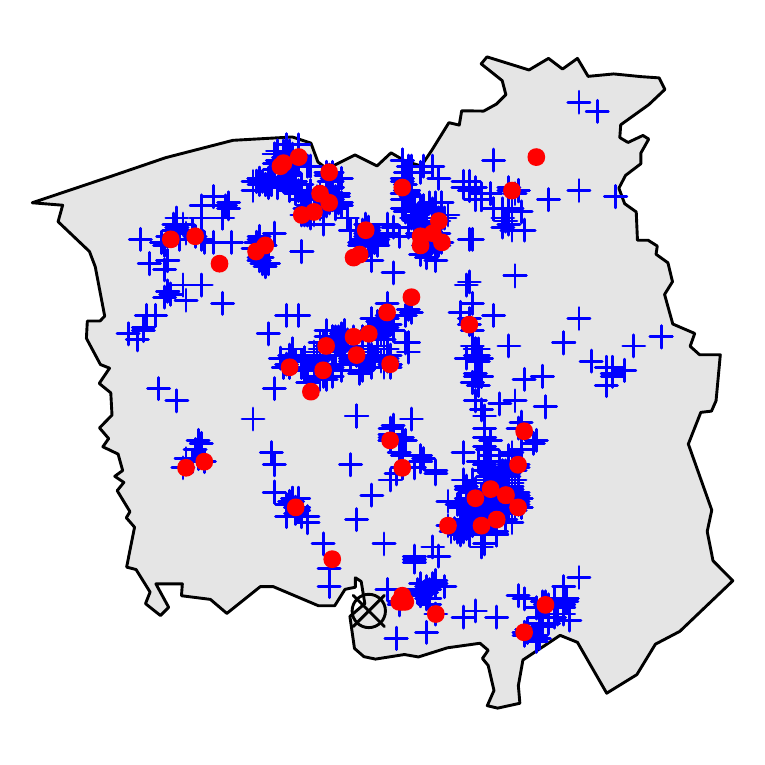}
\par\end{centering}
\caption{\label{fig:Locations-of-larynx}Locations of larynx (red dots) and
lung (blue pluses) cancers together with the location of the industrial
incinerator (black circle cross).}
\end{figure}

We apply the single pattern comparison test to a dataset from an epidemiological
study relating to the locations of larynx and lung cancer occurrences
in Chorley-Ribble area of Lancashire, England during the years of
1974-1983; the source of the data is \cite{DiggleIncinerator}. Figure
\ref{fig:Locations-of-larynx} plots the 58 cases of larynx cancer
and 978 cases of lung cancer together with the location of an industrial
incinerator in this area. Following \cite{DiggleIncinerator}, we
assume that the inhomogeneous Poisson process model applies, and that
the distribution of the lung cancer cases can be used as a surrogate
for the susceptible population. 

Let the corresponding true intensity functions be $\lambda^{\mathrm{Larynx}}(\cdot)$
and $\lambda^{\mathrm{Lung}}(\cdot)$. If there is an effect of the
relative location of the incinerator on larynx cancer (and not on
lung cancer), then the functional forms of these intensities would
be different, i.e. there would be some non-constant function $\rho(\cdot)$
such that $\lambda^{\mathrm{Larynx}}(\cdot)=\rho(\cdot)\lambda^{\mathrm{Lung}}(\cdot)$.
If, furthermore, $\rho(\cdot)$ depends on the distance to the incinerator,
this would be an evidence of a differential effect of the incinerator
on the two types of cancer.
\begin{figure}
\begin{centering}
\includegraphics[width=0.55\textwidth]{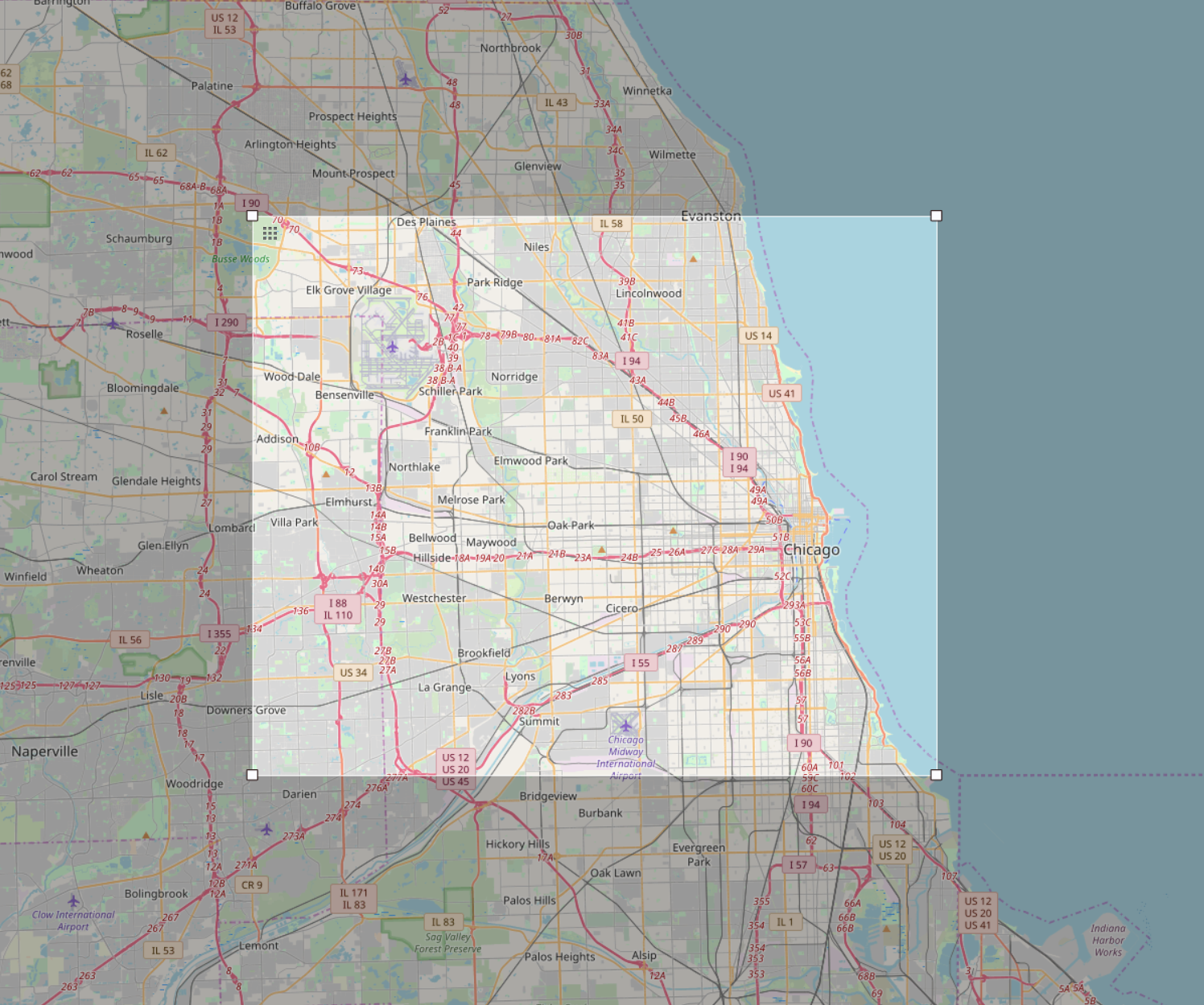}
\par\end{centering}
\caption{\label{fig:ChicagoBBOX}Geospatial window used for the Chicago Crime
experiment. The window coincides with the OpenStreetMap tile at zoom
level of 10 that covers the city of Chicago and the surrounding region.
Map data copyrighted OpenStreetMap contributors and available from
https://www.openstreetmap.org.}
\end{figure}

Considering the larynx cancer locations as the point pattern $\boldsymbol{X}$
and the lung cancer locations as the pattern $\boldsymbol{Y}$, we
would like to test whether the functional form of the intensity is
the same for these patterns. Note that the null hypothesis states
that there is a constant $c$ such that the intensities of the two
point processes satisfy $\lambda^{\mathrm{Larynx}}(\cdot)=c\lambda^{\mathrm{Lung}}(\cdot)$.
Applying our methodology, we find the $p$-value of $p^{H}=0.965$
and a mean Bayes factor of $\overline{\mathrm{BF}}_{10}=0.24$, which
leads us to retain the null hypothesis. This is in agreement with
the re-analyses of this data such as \cite{DigglePatComp1} and the
recent in-depth study of this dataset \cite{DAVIES201612} within
the context of difficulties that arise in estimating relative risk
functions.

\paragraph*{Replicated Patterns: Chicago Crime}

\setlength{\tabcolsep}{4pt}

\begin{table}
\begin{centering}
\begin{tabular}{lccccccc}
\toprule 
\multirow{2}{*}{Crime Type} & \multicolumn{1}{c}{Tuesday} & \multicolumn{1}{c}{Thursday} & \multicolumn{1}{c}{Saturday} & \multicolumn{2}{c}{Tue vs Thu} & \multicolumn{2}{c}{Tue vs Sat}\tabularnewline
\cmidrule{2-8} \cmidrule{3-8} \cmidrule{4-8} \cmidrule{5-8} \cmidrule{6-8} \cmidrule{7-8} \cmidrule{8-8} 
 & \#Pts & \#Pts & \#Pts & $p$-value & $\overline{\mathrm{BF}}_{10}$ & $p$-value & $\overline{\mathrm{BF}}_{10}$\tabularnewline
\midrule 
Theft & 144.7 & 150.3 & 148.6 & 0.125 & 0.838 & \textbf{0.001} & \textbf{29.503}\tabularnewline
Battery & 88.0 & 87.2 & 115.5 & 0.179 & 0.682 & \textbf{0.002} & \textbf{12.403}\tabularnewline
Deceptive Practice & 43.0 & 42.8 & 36.2 & 0.825 & 0.409 & \textbf{0.006} & \textbf{5.728}\tabularnewline
Criminal Damage & 50.5 & 53.6 & 61.4 & 0.566 & 0.502 & \textbf{0.014} & \textbf{2.799}\tabularnewline
Robbery & 19.9 & 19.4 & 22.9 & 0.099 & 0.858 & 0.051 & 1.231\tabularnewline
Narcotics & 28.1 & 27.1 & 30.1 & 0.915 & 0.392 & 0.613 & 0.486\tabularnewline
Burglary & 25.9 & 25.4 & 22.8 & 0.475 & 0.523 & 0.657 & 0.469\tabularnewline
Assault & 40.3 & 41.4 & 37.9 & 0.987 & 0.342 & 0.725 & 0.435\tabularnewline
Other Offense & 32.8 & 31.6 & 28.1 & 0.788 & 0.416 & 0.968 & 0.358\tabularnewline
Motor Veh Theft & 21.1 & 20.3 & 22.6 & 0.997 & 0.322 & 0.997 & 0.322\tabularnewline
\bottomrule
\end{tabular}
\par\end{centering}
\caption{\label{tab:chicago-crime}Application of the replicated pattern comparison
to Chicago Crime dataset. The entries in bold correspond to the rejected
hypotheses with the BH procedure at the FDR level of 0.1.}
\end{table}

\setlength{\tabcolsep}{\mytabcolsep}

In this experiment we use the dataset provided by Chicago Data Portal\footnote{data.cityofchicago.org}
that reflects reported incidents of crime that occurred in the City
of Chicago. We consider all of the crimes during the year of 2018
within a geospatial window (shown in Figure \ref{fig:ChicagoBBOX}),
and the goal is to compare the spatial patterns of each crime type
between weekdays and weekends. We wanted to pick days that are maximally
spaced out and avoid ambiguity (e.g. Friday nights), so we decided
to compare the following days of the week: Tuesday, Thursday, and
Saturday. The comparison between Tuesday and Thursday serves as a
sanity check, as we do not expect to see any differences between them. 

For each day and type of crime we obtain the crime patterns, and filter
by the count of points keeping the patterns that have at least 15
points. Next we count how many distinct patterns are available for
each day and type of crime; we found the crime types that had at least
40 patterns for each of Tuesday, Thursday, and Saturday and we kept
40 patterns per each to avoid differences in the testing power. The
types of crimes remaining after this filtering are shown in Table
\ref{tab:chicago-crime} together with the average count of points
per pattern.\emph{}

The results of comparisons are shown in the last two columns of Table
\ref{tab:chicago-crime}. We run the Benjamini-Hochberg \cite{FDR}
procedure on the 20 resulting $p$-values at the false discovery rate
of $0.1$, and the rejected hypotheses are indicated by the $p$-values
in bold. As expected, no differences were detected between Tuesday
and Thursday patterns. On the other hand, we see that there are statistically
significant differences between Tuesday and Saturday patterns in the
following categories of crime: theft, battery, deceptive practice,
and criminal damage. The corresponding mean Bayes factors can be interpreted
using Jeffreys\textquoteright{} rule of thumb \cite{Jeffreys61} for
strength of evidence pointing out that there is substantial or strong
evidence for the first three differences, whereas the difference in
criminal damage ``is not worth more than a bare mention''.

\section{Conclusion}

We have introduced an approach to detect differences in the first-order
structure of spatial point patterns. The proposed approach leverages
the kernel mean embedding in a novel way, by introducing its approximate
version. Hypothesis testing is based on conducting $t$-tests on each
dimension of the approximate embedding and combining them using either
the harmonic mean or Cauchy approach. Our experiments confirm that
the resulting tests are powerful and the $p$-values are well-calibrated.
Two applications to real world data have been presented. 

A number of possible extensions can be addressed in future research.
First, to focus the paper we avoided discussing hypothesis tests to
detect differences between more than two single patterns or more than
two groups of replicated patterns. Since aKME is based on means, under
suitable assumptions we can instead of $t$-tests perform an analysis
of variance (ANOVA) F test for differences among means and combine
the resulting $p$-values as before. Second, in the experimental section
we showed that the single pattern comparison test can be applied even
when Poisson assumption is not met; this was based on an intuitive
notion of effective sample size. This is different from the one discussed
in \cite{EffectiveSampleSize0,EffectiveSampleSize1,EffectiveSampleSize2},
and it would be interesting to formalize this notion and to provide
methods for its reliable estimation. Third avenue for future research
would be to leverage the directionality of the projections that correspond
to the subsets of dimensions of the aKME (c.f. Figure \ref{fig:rff_pic}).
One can envision a technique where anisotropy in the first-order structure
difference would be detected by appropriately combining the $p$-values.
An orthogonal aspect would be to investigate the scales at which the
differences happen, this time combining the $p$-values for each scale.
The harmonic mean approach is suitable for such multi-facet and multi-level
analyses as described in \cite{HarmonicP}. Fourth direction for future
work is studying the proposed approach in the context of goodness-of-fit
testing of the first-order structure. Since our approach allows for
vastly differing numbers of points in the patterns, one can compute
a low-variance approximation to the theoretical aKME of the model
distribution by constructing a pattern via drawing a large number
of points from that distribution; our initial studies gave favorable
results when compared to existing techniques such as \cite{YongtaoGuan2008}.
Finally, it would be desirable to extend the proposed technique to
the high-dimensional setting, where the curse of the dimensionality
would be an important factor that can affect the testing power. 

\paragraph*{Acknowledgements}

We are grateful to the revewers for their constructive comments which
have led to a much improved version of the article. We thank Alfred
Stein for his editorial efforts in handling this article for the Spatial
Statistics journal. Finally, we thank Tonglin Zhang and Isabel Fuentes-Santos
for providing the source code of the methods from their respective
papers.

\bibliographystyle{plain}
\addcontentsline{toc}{section}{\refname}\bibliography{biblio_els}
\appendix

\section{\label{sec:consistency}Proof of Consistency}

\setlength{\tabcolsep}{3pt}
\renewcommand{\arraystretch}{0.9}

\begin{table}
\begin{centering}
\begin{tabular}{lccccccccccccccc}
\toprule 
 &  &  &  &  &  &  & \multicolumn{3}{c}{{\footnotesize{}Harmonic}} &  &  &  & \multicolumn{3}{c}{{\footnotesize{}Adjusted Harmonic}}\tabularnewline
\midrule 
{\footnotesize{}Model} & {\footnotesize{}$\beta_{1}$} & {\footnotesize{}$\beta_{2}$} & {\footnotesize{}$\lambda$} &  &  &  & {\footnotesize{}$\alpha=0.01$} & {\footnotesize{}$\alpha=0.05$} & {\footnotesize{}$\alpha=0.1$} &  &  &  & {\footnotesize{}$\alpha=0.01$} & {\footnotesize{}$\alpha=0.05$} & {\footnotesize{}$\alpha=0.1$}\tabularnewline
\midrule
\midrule 
\textbf{\footnotesize{}Size} &  &  &  &  &  &  &  &  &  &  &  &  &  &  & \tabularnewline
{\footnotesize{}Linear} & {\footnotesize{}1} & {\footnotesize{}1} & {\footnotesize{}100} &  &  &  & {\footnotesize{}0.008} & {\footnotesize{}0.058} & {\footnotesize{}0.108} &  &  &  & {\footnotesize{}0.010} & {\footnotesize{}0.060} & {\footnotesize{}0.105}\tabularnewline
 &  & {\footnotesize{}1} & {\footnotesize{}400} &  &  &  & {\footnotesize{}0.009} & {\footnotesize{}0.045} & {\footnotesize{}0.098} &  &  &  & {\footnotesize{}0.007} & {\footnotesize{}0.038} & {\footnotesize{}0.085}\tabularnewline
 &  & {\footnotesize{}1} & {\footnotesize{}800} &  &  &  & {\footnotesize{}0.008} & {\footnotesize{}0.050} & {\footnotesize{}0.104} &  &  &  & {\footnotesize{}0.006} & {\footnotesize{}0.048} & {\footnotesize{}0.093}\tabularnewline
 & {\footnotesize{}2} & {\footnotesize{}2} & {\footnotesize{}100} &  &  &  & {\footnotesize{}0.011} & {\footnotesize{}0.054} & {\footnotesize{}0.106} &  &  &  & {\footnotesize{}0.013} & {\footnotesize{}0.048} & {\footnotesize{}0.104}\tabularnewline
 &  & {\footnotesize{}2} & {\footnotesize{}400} &  &  &  & {\footnotesize{}0.008} & {\footnotesize{}0.052} & {\footnotesize{}0.099} &  &  &  & {\footnotesize{}0.008} & {\footnotesize{}0.055} & {\footnotesize{}0.101}\tabularnewline
 &  & {\footnotesize{}2} & {\footnotesize{}800} &  &  &  & {\footnotesize{}0.010} & {\footnotesize{}0.044} & {\footnotesize{}0.090} &  &  &  & {\footnotesize{}0.008} & {\footnotesize{}0.042} & {\footnotesize{}0.094}\tabularnewline
 & {\footnotesize{}3} & {\footnotesize{}3} & {\footnotesize{}100} &  &  &  & {\footnotesize{}0.010} & {\footnotesize{}0.042} & {\footnotesize{}0.090} &  &  &  & {\footnotesize{}0.008} & {\footnotesize{}0.044} & {\footnotesize{}0.087}\tabularnewline
 &  & {\footnotesize{}3} & {\footnotesize{}400} &  &  &  & {\footnotesize{}0.012} & {\footnotesize{}0.058} & {\footnotesize{}0.106} &  &  &  & {\footnotesize{}0.010} & {\footnotesize{}0.049} & {\footnotesize{}0.104}\tabularnewline
 &  & {\footnotesize{}3} & {\footnotesize{}800} &  &  &  & {\footnotesize{}0.012} & {\footnotesize{}0.048} & {\footnotesize{}0.088} &  &  &  & {\footnotesize{}0.010} & {\footnotesize{}0.044} & {\footnotesize{}0.084}\tabularnewline
{\footnotesize{}Sine} & {\footnotesize{}1} & {\footnotesize{}1} & {\footnotesize{}100} &  &  &  & {\footnotesize{}0.010} & {\footnotesize{}0.054} & {\footnotesize{}0.100} &  &  &  & {\footnotesize{}0.008} & {\footnotesize{}0.058} & {\footnotesize{}0.100}\tabularnewline
 &  & {\footnotesize{}1} & {\footnotesize{}400} &  &  &  & {\footnotesize{}0.014} & {\footnotesize{}0.047} & {\footnotesize{}0.102} &  &  &  & {\footnotesize{}0.014} & {\footnotesize{}0.045} & {\footnotesize{}0.102}\tabularnewline
 &  & {\footnotesize{}1} & {\footnotesize{}800} &  &  &  & {\footnotesize{}0.012} & {\footnotesize{}0.059} & {\footnotesize{}0.106} &  &  &  & {\footnotesize{}0.008} & {\footnotesize{}0.058} & {\footnotesize{}0.112}\tabularnewline
 & {\footnotesize{}2} & {\footnotesize{}2} & {\footnotesize{}100} &  &  &  & {\footnotesize{}0.013} & {\footnotesize{}0.057} & {\footnotesize{}0.096} &  &  &  & {\footnotesize{}0.010} & {\footnotesize{}0.056} & {\footnotesize{}0.102}\tabularnewline
 &  & {\footnotesize{}2} & {\footnotesize{}400} &  &  &  & {\footnotesize{}0.014} & {\footnotesize{}0.058} & {\footnotesize{}0.094} &  &  &  & {\footnotesize{}0.012} & {\footnotesize{}0.056} & {\footnotesize{}0.093}\tabularnewline
 &  & {\footnotesize{}2} & {\footnotesize{}800} &  &  &  & {\footnotesize{}0.008} & {\footnotesize{}0.046} & {\footnotesize{}0.095} &  &  &  & {\footnotesize{}0.007} & {\footnotesize{}0.042} & {\footnotesize{}0.094}\tabularnewline
 & {\footnotesize{}3} & {\footnotesize{}3} & {\footnotesize{}100} &  &  &  & {\footnotesize{}0.008} & {\footnotesize{}0.044} & {\footnotesize{}0.092} &  &  &  & {\footnotesize{}0.010} & {\footnotesize{}0.050} & {\footnotesize{}0.098}\tabularnewline
 &  & {\footnotesize{}3} & {\footnotesize{}400} &  &  &  & {\footnotesize{}0.008} & {\footnotesize{}0.051} & {\footnotesize{}0.090} &  &  &  & {\footnotesize{}0.006} & {\footnotesize{}0.046} & {\footnotesize{}0.093}\tabularnewline
 &  & {\footnotesize{}3} & {\footnotesize{}800} &  &  &  & {\footnotesize{}0.010} & {\footnotesize{}0.046} & {\footnotesize{}0.091} &  &  &  & {\footnotesize{}0.012} & {\footnotesize{}0.048} & {\footnotesize{}0.091}\tabularnewline
\midrule 
\textbf{\footnotesize{}Power} &  &  &  &  &  &  &  &  &  &  &  &  &  &  & \tabularnewline
{\footnotesize{}Linear} & {\footnotesize{}1} & {\footnotesize{}2} & {\footnotesize{}100} &  &  &  & {\footnotesize{}0.085} & {\footnotesize{}0.206} & {\footnotesize{}0.310} &  &  &  & {\footnotesize{}0.086} & {\footnotesize{}0.198} & {\footnotesize{}0.300}\tabularnewline
 &  & {\footnotesize{}2} & {\footnotesize{}400} &  &  &  & {\footnotesize{}0.644} & {\footnotesize{}0.816} & {\footnotesize{}0.882} &  &  &  & {\footnotesize{}0.627} & {\footnotesize{}0.818} & {\footnotesize{}0.879}\tabularnewline
 &  & {\footnotesize{}2} & {\footnotesize{}800} &  &  &  & {\footnotesize{}0.972} & {\footnotesize{}0.992} & {\footnotesize{}0.996} &  &  &  & {\footnotesize{}0.962} & {\footnotesize{}0.990} & {\footnotesize{}0.995}\tabularnewline
 &  & {\footnotesize{}3} & {\footnotesize{}100} &  &  &  & {\footnotesize{}0.587} & {\footnotesize{}0.762} & {\footnotesize{}0.848} &  &  &  & {\footnotesize{}0.538} & {\footnotesize{}0.756} & {\footnotesize{}0.852}\tabularnewline
 &  & {\footnotesize{}3} & {\footnotesize{}400} &  &  &  & {\footnotesize{}1.000} & {\footnotesize{}1.000} & {\footnotesize{}1.000} &  &  &  & {\footnotesize{}1.000} & {\footnotesize{}1.000} & {\footnotesize{}1.000}\tabularnewline
 &  & {\footnotesize{}3} & {\footnotesize{}800} &  &  &  & {\footnotesize{}1.000} & {\footnotesize{}1.000} & {\footnotesize{}1.000} &  &  &  & {\footnotesize{}1.000} & {\footnotesize{}1.000} & {\footnotesize{}1.000}\tabularnewline
 & {\footnotesize{}2} & {\footnotesize{}3} & {\footnotesize{}100} &  &  &  & {\footnotesize{}0.070} & {\footnotesize{}0.179} & {\footnotesize{}0.274} &  &  &  & {\footnotesize{}0.058} & {\footnotesize{}0.182} & {\footnotesize{}0.270}\tabularnewline
 &  & {\footnotesize{}3} & {\footnotesize{}400} &  &  &  & {\footnotesize{}0.520} & {\footnotesize{}0.725} & {\footnotesize{}0.808} &  &  &  & {\footnotesize{}0.496} & {\footnotesize{}0.718} & {\footnotesize{}0.815}\tabularnewline
 &  & {\footnotesize{}3} & {\footnotesize{}800} &  &  &  & {\footnotesize{}0.915} & {\footnotesize{}0.968} & {\footnotesize{}0.982} &  &  &  & {\footnotesize{}0.912} & {\footnotesize{}0.968} & {\footnotesize{}0.981}\tabularnewline
{\footnotesize{}Sine} & {\footnotesize{}1} & {\footnotesize{}2} & {\footnotesize{}100} &  &  &  & {\footnotesize{}0.352} & {\footnotesize{}0.608} & {\footnotesize{}0.720} &  &  &  & {\footnotesize{}0.338} & {\footnotesize{}0.564} & {\footnotesize{}0.710}\tabularnewline
 &  & {\footnotesize{}2} & {\footnotesize{}400} &  &  &  & {\footnotesize{}0.997} & {\footnotesize{}0.999} & {\footnotesize{}1.000} &  &  &  & {\footnotesize{}0.998} & {\footnotesize{}0.999} & {\footnotesize{}1.000}\tabularnewline
 &  & {\footnotesize{}2} & {\footnotesize{}800} &  &  &  & {\footnotesize{}1.000} & {\footnotesize{}1.000} & {\footnotesize{}1.000} &  &  &  & {\footnotesize{}1.000} & {\footnotesize{}1.000} & {\footnotesize{}1.000}\tabularnewline
 &  & {\footnotesize{}3} & {\footnotesize{}100} &  &  &  & {\footnotesize{}0.942} & {\footnotesize{}0.985} & {\footnotesize{}0.990} &  &  &  & {\footnotesize{}0.940} & {\footnotesize{}0.982} & {\footnotesize{}0.992}\tabularnewline
 &  & {\footnotesize{}3} & {\footnotesize{}400} &  &  &  & {\footnotesize{}1.000} & {\footnotesize{}1.000} & {\footnotesize{}1.000} &  &  &  & {\footnotesize{}1.000} & {\footnotesize{}1.000} & {\footnotesize{}1.000}\tabularnewline
 &  & {\footnotesize{}3} & {\footnotesize{}800} &  &  &  & {\footnotesize{}1.000} & {\footnotesize{}1.000} & {\footnotesize{}1.000} &  &  &  & {\footnotesize{}1.000} & {\footnotesize{}1.000} & {\footnotesize{}1.000}\tabularnewline
 & {\footnotesize{}2} & {\footnotesize{}3} & {\footnotesize{}100} &  &  &  & {\footnotesize{}0.050} & {\footnotesize{}0.190} & {\footnotesize{}0.308} &  &  &  & {\footnotesize{}0.034} & {\footnotesize{}0.144} & {\footnotesize{}0.256}\tabularnewline
 &  & {\footnotesize{}3} & {\footnotesize{}400} &  &  &  & {\footnotesize{}0.768} & {\footnotesize{}0.912} & {\footnotesize{}0.942} &  &  &  & {\footnotesize{}0.784} & {\footnotesize{}0.912} & {\footnotesize{}0.943}\tabularnewline
 &  & {\footnotesize{}3} & {\footnotesize{}800} &  &  &  & {\footnotesize{}0.994} & {\footnotesize{}1.000} & {\footnotesize{}1.000} &  &  &  & {\footnotesize{}0.994} & {\footnotesize{}0.999} & {\footnotesize{}1.000}\tabularnewline
\bottomrule
\end{tabular}
\par\end{centering}
\caption{\label{tab:harmonic-vs-adjusted}Comparison of rejection rates for
the single pattern comparison test between harmonic and adjusted harmonic
approaches.}
\end{table}

\setlength{\tabcolsep}{\mytabcolsep}
\renewcommand{\arraystretch}{1.0}

Here we provide the proof of consistency of the proposed test for
single pattern comparison with increasing number of points in the
patterns. A similar argument applies to the replicated pattern comparison
test with the increasing number of replications; this, of course,
assumes that the condition of Eq. (\ref{eq:condition}) holds. 

Consider two density functions $p(\cdot)\neq q(\cdot)$ and the corresponding
inhomogeneous Poisson point processes $P$ and $Q$ such that $\lambda^{P}(\cdot)=Np(\cdot)$
and $\lambda^{Q}(\cdot)=Nq(\cdot)$. We will first show that there
exists $D=2m_{0}\ell_{0}$ dimensional aKME embedding based test such
that for any size $\alpha$, we have $\mathrm{Prob}(\mathrm{reject}\,H_{0})\to1$
when $N\to\infty$. Afterwards, we will discuss how to  modify our
test and slowly increase $D$ with $N$ so as to obtain general consistency. 

As discussed in Section \ref{sec:Approximate-Kernel-Mean}, we have
the following connection to MMD,
\[
\Vert\mu_{m,\ell}^{p}-\mu_{m,\ell}^{q}\Vert^{2}\to MMD^{2}(p,q),
\]
with increasing quality of quadrature approximation as $m,\ell\to\infty$.
For a large class of density functions, $p(\cdot)\neq q(\cdot)$ implies
$MMD^{2}(p,q)>0$ \cite{mmd}. As a result, there exist $m_{0}$ and
$\ell_{0}$ such that the $D=2m_{0}\ell_{0}$ dimensional aKME satisfies
$\Vert\mu_{m_{0},\ell_{0}}^{p}-\mu_{m_{0},\ell_{0}}^{q}\Vert^{2}>0$.
Thus, there is a dimension $k$ in the embedding space such that $1\leq k\leq D$
and aKME embeddings disagree on this dimension: $(\mu_{m_{0},\ell_{0}}^{p})_{k}\neq(\mu_{m_{0},\ell_{0}}^{q})_{k}$.
Note that the value of $D$ depends only on $p$ and $q$, but not
on $N$.

Of course, the difference in the $k$-th dimension will be discerned
by the $t$-test as $N\to\infty$: we will have the corresponding
$p$-value $p_{k}\to0$. More precisely, for any $\delta>0$, $\mathrm{Prob}(p_{k}<\delta)\to1.$
If we were to use Bonferroni correction instead of combining $p$-values,
we would get for a test size $\alpha,$ the following $\mathrm{Prob}(\mathrm{reject}\,H_{0})=\mathrm{Prob}(p_{k}<\alpha/D)\to1$
as $N\to\infty$. We will see that both harmonic mean combination
and Cauchy combination tests are asymptotically equivalent to Bonferroni
correction in the worst-case regime, yielding consistency.

We first prove consistency for the $p^{H}$ based test. In addition
to $p_{k}$ we have $D-1$ other $p$-values coming from testing the
remaining coordinates of the embedding. To consider the worst-case
scenario, assume that the null holds for all of these $D-1$ coordinates.
For the harmonic mean combination, the combined $p$-value can be
upper-bounded by substituting the null $p$-values by 1. Plugging
in $p_{i}=1,i\neq k$ we get, 
\[
p^{H}\leq H\left(\frac{D}{D-1+\frac{1}{p_{k}}}\right)=H\left(\frac{Dp_{k}}{1+(D-1)p_{k}}\right)\sim H(Dp_{k})\sim Dp_{k},
\]
for small $p_{k}$; here we used $\lim_{x\to0}H(x)/x\to1$. This proves
that as $N\to\infty$, $\mathrm{Prob}(\mathrm{reject}\,H_{0})=\mathrm{Prob}(p^{H}<\alpha)\geq\mathrm{Prob}(Dp_{k}<\alpha)=\mathrm{Prob}(p_{k}<\alpha/D)\to1$
for any level $\alpha>0$. 

To carry out a similar limit argument for the Cauchy combination test,
more care is needed to avoid the singularity around $p$-values close
to 1. Using approximation by normal distribution and the commonly
used bounds on the tail area, one can show that for the non-null $t$-test
the $p$-values is of the order $p_{k}\sim e^{-\theta N}/\sqrt{N}$
for some $\theta$ that depends on the effect size; so, we have $\mathrm{Prob}(p_{k}<e^{-\theta N})\to1$.
Take an arbitrary sequence $\epsilon_{N}\to0$ such that the convergence
happens slower than $p_{k}\to0$ with $N\to\infty$, e.g. $\epsilon_{N}=1/N^{r}$
for any positive $r$ would work. Considering the worst-case scenario
above, we have that with probability $(1-\epsilon_{N})^{D-1}$ the
following holds: $p_{i}\leq1-\epsilon_{N}$ $\forall i\neq k$; this
stems from the uniform distribution of $p$-values under the null.
Now to obtain an upper bound for the Cauchy combination $p$-value
$p^{C}$ we can replace all of the null $p$-values with $1-\epsilon_{N}$,
yielding with at least probability of $(1-\epsilon_{N})^{D-1}$:
\[
\cot\pi p^{C}\geq\frac{(D-1)\cot\pi(1-\epsilon_{N})+\cot\pi p_{k}}{D}\sim\frac{1}{D\pi}(p_{k}^{-1}-(D-1)\epsilon_{N}^{-1})\sim\frac{1}{D\pi}e^{\theta N},
\]
using simple asymptotics for the cotangent function and by the assumption
on $\epsilon_{N}$. The inequality implies smallness of $p^{C}$which
allows us to use the asymptotic $\cot x\sim1/x$ once again to get
\[
\mathrm{Prob}(p^{C}\leq De^{-\theta N})\geq(1-\epsilon_{N})^{D-1},
\]
in other words, Cauchy combination test is asymptotically equivalent
to Bonferroni correction in this regime. With $N\to\infty$ we have
$p^{C}\to0$ with probability $(1-\epsilon_{N})^{D-1}\to1$, proving
consistency for the Cauchy version of our test. 

The above argument assumes the knowledge of $D=m_{0}\ell_{0}$ when
conducting the test. To avoid this assumption, we need to increase
the dimensionality of aKME with $N$. To this end, consider two sequences
of integers that satisfy $m_{N}\to\infty$ and $\ell_{N}\to\infty$
as $N\to\infty$, potentially with repeating terms (e.g. they can
sweep all the pairwise combinations of integers $m$ and $\ell$).
We define $\mathrm{aKME}_{N}$ as the concatenation of all \textit{unique}
$\mathrm{aKME}_{m_{s},\ell_{s}}$, $s\leq N$ and apply the $t$-tests
to each dimension of $\mathrm{aKME}_{N}$. Denote the dimensionality
of this overall embedding as $D_{N}$. As $N\to\infty$, we are guaranteed
to include a non-null dimension into the overall $\mathrm{aKME}_{N}$
and it will stay in due to the nestedness of this construction. As
explained above we can use Bonferroni correction as a surrogate for
both of the combination techniques. Thus, concentrating on Bonferroni
correction, we have $\mathrm{Prob}(\mathrm{reject}\,H_{0})=\mathrm{Prob}(p_{k}<\alpha/D_{N})$.
Since $\mathrm{Prob}(p_{k}<e^{-\theta N})\to1$, we see that if, for
example, $D_{N}$ grows polynomially (which is easy to attain) consistency
will follow. For the Cauchy combination test one has to be slightly
more careful and select $\epsilon_{N}$ in the previous paragraph
that goes to zero faster than, say, $1/D_{N}^{2}$.

\section{\label{sec:R-Implementation}Implementation of Single Pattern Comparison
in R}

Here we provide an R implementation of the single pattern comparison
test for the example demonstrated in Section \ref{subsec:Real-World-Data},
namely comparison of the larynx and lung cancer location patterns.

\begin{lstlisting}[language=R, caption={}]
library(spatstat)
library(harmonicmeanp)
library(BayesFactor)

radialGaussHermiteData <- function(k) {
	if (k == 4) {
		Roots4 = c(
			0.3961205684809970482046989062,
			1.1767346714185921750900701015,
			2.2010676629189581946543092946,
			3.4836109980286615716953100856
		)
		Weights4 = c(
			0.2279981086730596270303131898,
			0.538464759426994159094177725,
			0.2215779133653168055615931299,
			0.0119592185346294083139159552
		)
	    list(x = Roots4, w = Weights4)
	} else {
		stop(
			"Please refer to the values in
			http://www.jaeckel.org/RootsAndWeightsForRadialGaussHermiteQuadrature.cpp"
		)
	} 
}

#maps every point in the point pattern using aKME process
get_aKME <- function(p, sigmas, n_radial = 4, n_linear = 4) {
	#radial unit vectors
	theta = seq(0, pi, length.out = n_radial + 1)[-(n_radial + 1)]
	W0 = cbind(cos(theta), sin(theta))
	
	#quadrature roots and weights
	temp = radialGaussHermiteData(n_linear)
	Roots = temp$x
	Weights = temp$w
	
	W = do.call('rbind', lapply(seq_along(Roots), function(i) Roots[i] * W0))
	coeff = rep(sqrt(Weights / n_radial), each=nrow(W0))
	
	#project the points onto the lines, c.f. Figure 1
	proj = W%*%rbind(p$x, p$y)
	
	#embedding: for each sigma wrap on the circle of corresponding radius
	emb = lapply(seq_along(sigmas), function(i)
		t(rbind(cos(proj / sigmas[i]) * coeff, sin(proj / sigmas[i]) * coeff)))
	
	#concatenated embedding, one row for each point of the pattern
	#dim = (#points,  #sigmas x 2 x # n_radial x n_linear)
	do.call('cbind', emb)
}

#load data
d = chorley
d1 = split(d)$lung
d2 = split(d)$larynx

#we use several Gaussian widths at once
#these values are for [0,1]x[0,1] window
sigma_list0 = c(1 / 16, 1 / 8, 1 / 4) 
sigma_list = diameter(Window(d)) * sigma_list0 / sqrt(2)

aKME1 = get_aKME(d1, sigma_list)
aKME2 = get_aKME(d2, sigma_list)

#p-values for t-tests on each dimension of the embeddings
p_ttests = sapply(seq_len(ncol(aKME1)), function(i)
	t.test(aKME1[, i], aKME2[, i])$p.value)

#overall p-value using harmonic mean combo
p_harmonic = p.hmp(p_ttests, L = length(p_ttests), multilevel=FALSE)
#0.9654329

#overall p-value using Cauchy combo
p_cauchy = 0.5 - atan(mean(tan((0.5-p_ttests)*pi)))/pi
#0.6727789

#mean Bayes factor
bf10_mean = mean(sapply(seq_len(ncol(aKME1)), function(i)
	extractBF(ttestBF(aKME1[, i], aKME2[, i]))$bf))
#0.2443478
\end{lstlisting}
\end{document}